\documentclass{article}

\usepackage{PRIMEarxiv}

\usepackage{booktabs} 
\usepackage{authblk}
\usepackage{multirow}
\usepackage{amssymb}
\usepackage{amsmath}
\usepackage{algpseudocode}
\usepackage{algorithm}
\usepackage{graphicx} 
\usepackage{subcaption} 
\usepackage[table]{xcolor} 
\usepackage{mathrsfs} 

\title{WeiDetect: Weibull Distribution-Based Defense against Poisoning Attacks in Federated Learning for Network Intrusion Detection Systems}

\author[1]{Sameera K. M.}
\author[2]{Vinod P.\thanks{Corresponding author: vinod.puthuvath@unipd.it}}
\author[3]{Anderson Rocha}
\author[1]{Rafidha Rehiman K. A.}
\author[2]{Mauro Conti}

\affil[1]{Department of Computer Applications, Cochin University of Science and Technology, India}
\affil[2]{Department of Mathematics, University of Padua, Italy}
\affil[3]{University of Campinas (Unicamp), Campinas, Brazil}

\affil[ ]{\texttt{sameerakm@cusat.ac.in, vinod.p@cusat.ac.in, vinod.puthuvath@unipd.it}}
\affil[ ]{\texttt{arrocha@unicamp.br, rafidharehimanka@cusat.ac.in, mauro.conti@unipd.it}}

\begin{document}
\maketitle
\begin{abstract}
In the era of data expansion, ensuring data privacy has become increasingly critical, posing significant challenges to traditional AI-based applications. In addition, the increasing adoption of IoT devices has introduced significant cybersecurity challenges, making traditional Network Intrusion Detection Systems (NIDS) less effective against evolving threats, and privacy concerns and regulatory restrictions limit their deployment. Federated Learning (FL) has emerged as a promising solution, allowing decentralized model training while maintaining data privacy to solve these issues. However, despite implementing privacy-preserving technologies, FL systems remain vulnerable to adversarial attacks. Furthermore, data distribution among clients is not heterogeneous in the FL scenario. We propose WeiDetect, a two-phase, server-side defense mechanism for FL-based NIDS that detects malicious participants to address these challenges. In the first phase, local models are evaluated using a validation dataset to generate validation scores. These scores are then analyzed using a Weibull distribution, identifying and removing malicious models. We conducted experiments to evaluate the effectiveness of our approach in diverse attack settings. Our evaluation included two popular datasets, CIC-Darknet2020 and CSE-CIC-IDS2018, tested under non-IID data distributions. Our findings highlight that WeiDetect outperforms state-of-the-art defense approaches, improving higher target class recall up to 70\% and enhancing the global model's F1 score by 1\% to 14\%. 

\end{abstract}












\section{Introduction}
\label{sec:introduction}

The rapid advancement of the Internet has created a highly interconnected world. The adoption of IoT for connectivity has increased significantly, leading to security vulnerabilities due to the inherent nature of IoT devices and systems. According to \cite{THEIN2024182}, it was emphasized that IoT devices are expected to reach 55.7 billion by 2025; the increasing volume of data generated by these devices also opens the door to cyber attackers. This further signifies the critical role of the Network Intrusion Detection System~(NIDS), which detects suspicious activities and improves the security of the IoT network ecosystem. 
\par 
The NIDS employs signature, behavior, or specification-based approaches to identify network anomalies and protect the system from unauthorized use or access~\cite{LIAO201316}. However, these approaches have become less efficient in recognizing unknown attacks, rendering them incapable of detecting new or evolving threats. The paper
 \cite{ALHAWAWREH20181, khan2021enhanced} highlight that Machine Learning~(ML) based NIDSs are efficient alternatives that identify normal and abnormal traffic patterns in IoT systems. Although these ML models have been widely employed in various solutions to enable dynamic and adaptive IDS in IoT environments. However, they often rely on a central entity to accumulate data from all IoT devices, raising significant concerns about data privacy in the central entity during the learning process. Furthermore, ML-based approaches may not be significant for large-scale networks due to their computational and resource-intensive nature~\cite {yang2019federated}. The centralized processing of various 5G or 6G data, including text, audio, and video, is time-consuming and costly, especially in IDSs.
\par 
To address the abovementioned concerns, researchers in \cite{mcmahan17a} proposed a distributed learning paradigm called Federated Learning (FL). FL facilitates multiple IoT devices to collaboratively train a shared model without directly sharing their private data with a central entity, ensuring compliance with current legal regulations. This approach also facilitates AI benefits to heterogeneous domains that handle sensitive data and enable a large-scale training process, particularly FL-based NIDS~\cite{FRIHA202217,SU2024}.  
 \par 
Despite its benefits, FL-based NIDS remains vulnerable to security threats, particularly poisoning attacks from malevolent user devices. Furthermore, the central entity responsible for the aggregation lacks direct access to local training datasets. As a result, it cannot effectively monitor malicious activities within the system or determine the credibility of the data. In real-world circumstances, fluctuations in traffic patterns, attack behaviors, and device characteristics lead to a significant variation in the client's data distribution. This divergence complicates the learning process and weakens the model’s ability to generalize across diverse network environments, making it more susceptible to adversarial manipulations \cite{karimireddy2020scaffold}.

Moreover, existing robust aggregation mechanisms struggle to mitigate the impact of poisoned clients as data heterogeneity increases, ultimately degrading the performance of FL-based NIDS. Even without poisoned clients, imbalanced data distributions can negatively affect model performance. Since FL relies on decentralized data to enhance a unified global model, variations in client data distribution due to heterogeneity can hinder convergence. In highly heterogeneous environments, local datasets differ significantly from the global distribution, causing local training objectives to diverge from the global optimum. Consequently, local updates may drift away from the global objective, particularly when models undergo extensive local training (e.g., with many local epochs). As a result, the final global model achieves significantly lower accuracy than an IID setting.

\begin{figure}[h!]
    \centering
    \includegraphics[width=\linewidth]{ 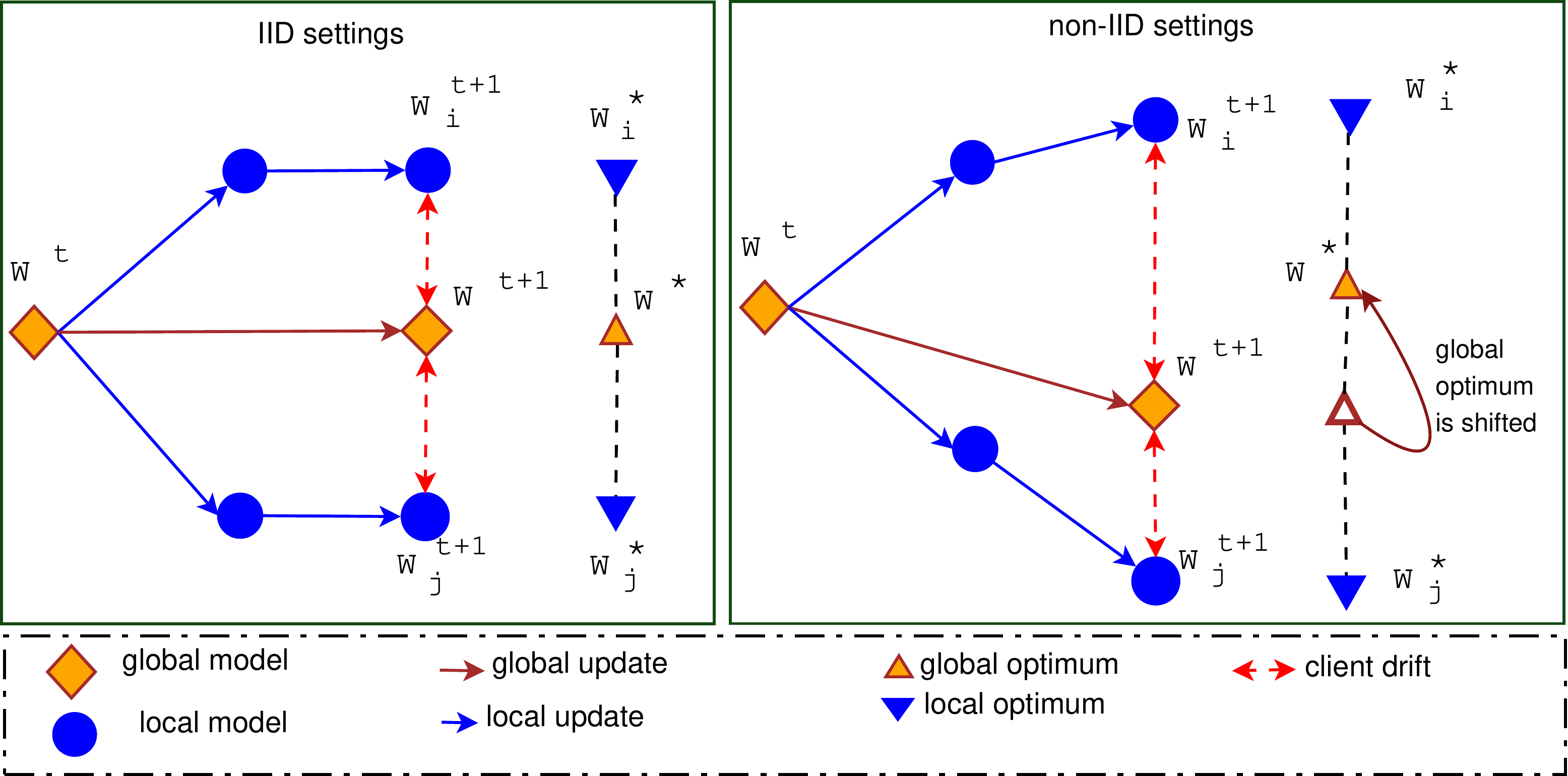}
    \caption{Visualization of client drift in IID and non-IID data distributions.}
    \label{fig:client_drift}
\end{figure}

Figure \ref{fig:client_drift} visualizes the convergence of the global model in the scenarios of IID and non-IID data using the FedAvg approach. In the IID setting, the averaged global optimum $ w^*$ is located at the center of the local optima of both clients $ w_i^*$ and $ w_j^*$, ensuring stable convergence. However, in the non-IID setting, client drift due to an imbalanced data distribution causes $ w^*$ to shift away from the true global optimum, preventing the global model from converging effectively.

\par Researchers have shown that adversaries can exploit compromised devices to manipulate local model updates by data poisoning or model poisoning techniques \cite{nguyen2020poisoning,MOTHUKURI2021619}. Data poisoning involves modifying the training data by flipping labels or injecting adversarial samples, while model poisoning directly alters the model parameters to embed malicious behaviors. Label flipping involves altering ground truth labels without changing the traffic features. At the same time, adversarial samples are meticulously crafted inputs designed to mislead the model by making subtle, imperceptible changes to legitimate data. With these techniques, an adversary aims to degrade the performance of the NIDS and, consequently, the network. For example, by poisoning the data, FL-based NIDS may misclassify benign traffic as malicious or malicious traffic as benign~\cite{agrawal2022federated}. 

Based on the attacker's objective, data poisoning is further categorized into targeted or untargeted, where targeted attacks manipulate specific classes, while untargeted attacks aim to degrade overall model performance. Additionally, adversaries may construct deceptive models by leveraging corrupted data, further compromising the integrity of the global model. ~\cite{goodfellow2014explaining,madry2017towards} introduce adversarial sample generation techniques that assume white-box access to the training model. By injecting controlled perturbations and solving an optimization problem, they generate adversarial instances designed to degrade the global model performance. Our work motivates us to craft deterministic and targeted adversarial instances specifically for compromised participants. 
\par Various defense approaches have been introduced to mitigate the impact of these poisoning strategies~\cite{ZHANG2022154,taheri2020fed,JEBREEL2023110178,fung1808mitigating,shejwalkar2021manipulating}. These approaches face significant challenges: First, they often need to identify the intricate patterns required to effectively detect poisoned models, especially when dealing with the high dimensionality of learning parameters. Secondly, the heterogeneous and time-varying nature of IoT traffic data results in non-IID distributions, leading to significant shifts between pre-trained and real-time local models and sabotaging global model accuracy. Additionally, class imbalance in local datasets further exacerbates performance degradation, as non-robust aggregation algorithms fail to generalize effectively. The lack of adequate validation across distributed devices also amplifies these challenges, ultimately impacting the reliability of the global model. 
In IoT environments, untrustworthy collaborators in federated training introduce significant security risks, especially when data distributions vary among participants. While existing approaches, such as verifying model weights or gradients, aim to enhance security, they remain insufficient for detecting data poisoning attacks effectively. Therefore, further research is essential to comprehensively analyze these threats and develop robust defense mechanisms for FL-based NIDS. 

In this research, we propose a Weibull distribution-based defense approach, WeiDetect, to address these challenges and ensure the robustness of FL-based NIDS against data poisoning attacks. The Weibull distribution~\cite{ndeba2025comparative} provides a probabilistic framework for modeling and analyzing outlier behaviors, making it suitable for detecting attack patterns in non-IID FL scenarios. Its flexibility in handling data sparsity, 
adaptability to distribution shifts, and effectiveness in identifying anomalous patterns made Weibull distribution suitable for non-IID data distributions. This helps in accurately detecting compromised or low-quality models, effectively overcoming the limitations of current approaches. The authors in\cite{R2020262} employ the Weibull distribution to detect faulty data and measure data quality in cyber-physical systems. Their result shows that the approach improved detection accuracy and adaptability in heterogeneous and dynamic IoT environments. WeiDetect is a two-phase defense strategy designed to enhance the robustness of NIDS systems against data poisoning attacks. In the first phase, local models are evaluated against a validation dataset to generate performance scores. These scores are then analyzed using a Weibull distribution. The Cumulative Distribution Function (CDF) is calculated to identify and eliminate outlier models indicative of poisoned or unreliable models. By removing these outliers, WeiDetect ensures that only benign models contribute to the global model. We have conducted various experiments with  CIC-Darknet2020, and CSE-CIC-IDS2018 datasets in non-IID scenarios under different attack settings. Additionally, we compared the performance of WeiDetect with other state-of-the-art approaches. 
The key contributions of our study are given below.
\begin{itemize}
    \item We frame targeted data poisoning attacks in FL-based NIDS by generating adversarial samples. Specifically, we adapt various attack algorithms to craft these adversarial samples. 
    \item We propose WeiDetect, a robust FL-based NIDS defense strategy for the heterogeneous environment utilizing the Weibull distribution approach to detect malicious devices. WeiDetect leverages a two-phase defense mechanism to safeguard the global intrusion detection model from the impact of malicious nodes. 
    \item We evaluate the performance of WeiDetect using the CIC-Darknet2020 and CSE-CIC-IDS2018 datasets, simulating imbalanced data and experimenting with targeted poisoning attacks at various levels. Furthermore, we compared WeiDetect with the existing state-of-the-art defense mechanisms to evaluate its performance.
    \item We additionally evaluate WeiDetect's performance against Label Flipping attacks, extending the evaluation to single-label and multi-label attack scenarios.
\end{itemize}
\par  \textbf{Roadmap.} This paper is structured as follows. Section \ref{sec:backgroundrelatedwork} reviews related work in the field. Section \ref{sec:systemmodel} presents the system model, and Section \ref{sec:proposedmethod} outlines the proposed WeiDetect. Section \ref{sec:experimentalsection} and \ref{sec:discussion} detailed results and discussion. Finally, Section \ref{sec:conclusion} concludes WeiDetect with future research directions.
\section{Background and Related Work}
\label{sec:backgroundrelatedwork}

\subsection{FL-based network intrusion detection system}
\label{subsec1}
FL has emerged as a new avenue for promising privacy-preserving solutions widely applied in NIDS for IoT networks~\cite{AMIRIZARANDI2023110005}. FL ensures low communication costs and low latency without compromising the accuracy of the global model. Friha et al. proposed \cite{FRIHA202217} FELIDS, which uses a deep learning approach to mitigate cyberattacks in agricultural IoT infrastructures. 
Li et al.~\cite{LI9195012} developed DeepFed, an intrusion detection system for the Industrial Internet of Things (IIoT), employing CNN and gated recurrent units for effective local feature extraction and FL for secure data aggregation.

\par Researchers in \cite{Fan9343358} proposed a customized NIDS leveraging federated transfer learning to improve performance. Nguyen et al. \cite{Nguyen8884802} developed DÏOT, an autonomous system for detecting compromised IoT devices without relying on labeled data or human input. It leverages device-specific communication profiles and FL to identify unusual behavior and adapt to emerging threats. DÏOT demonstrated a $95.6\%$ detection rate with minimal latency, approximately $257 ms$. However, it is limited to detecting attacks on IoT devices and does not cover threats to other network components - SDN or services and protocols such as FTP and SSH. In \cite{Mothukuri9424138}, the authors proposed an FL-based intrusion detection model that enables resource-constrained devices to share local knowledge without exposing data content. The proposed approach reaches an accuracy closer to the centralized approach. However, this research does not further analyze their robustness against poisoning attacks. The paper \cite{TABASSUM2022299} also proposed privacy-preserving FL-based IDS, which used a Generative Adversarial Network (GAN) based model to secure the network with high accuracy.

\subsection{Poisoning attack in FL-based NIDS}

\label{sec:Rpoisoningattack}
FL-based NIDS face significant vulnerabilities from data poisoning attacks, where adversaries inject malicious or mislabeled data into local training datasets. In data poisoning attacks, the adversary modifies the features of the samples or alters their labels to impair the performance of the intrusion detection model. Some researchers are studying the impact of data poisoning attacks on FL-based NIDS. For instance, \cite{nguyen2020poisoning} shows that FL-based NIDS are susceptible to backdoor attacks, where adversaries implant backdoors in the aggregated model, leading to misclassifying malicious traffic as benign. These covert attacks leverage adversarially crafted malicious samples from compromised IoT devices and remain undetected during training. Evaluations of real-world datasets from 46 devices demonstrated the effectiveness of this strategy in bypassing advanced FL defenses. In another study \cite{Zhang10074658}, authors created a clean-label data poisoning attack using a GAN called PT-GAN, incorporating feedback from the target model. PT-GAN generates minimally perturbed, correctly labeled traffic samples injected into the local training dataset to corrupt the intrusion detection model. The result demonstrated that FL-based NIDS degraded performance by 28\% under PT-GAN attack.
\par Authors in \cite{ZHANG2022154} examine poisoning attacks on FL-based NIDS, including Label Flipping and clean-label attacks. They use the C\&W algorithm to generate adversarial samples for the clean-label attack. Similarly, \cite{THEIN2024182} analyzes label-flipping and model poisoning attacks on FL-based IDS. The gap in this research is that it does not address the combined impact of different data poisoning attacks on the model's performance.
\begin{figure*}[h!]
    \centering
    \includegraphics[width=0.8\linewidth]{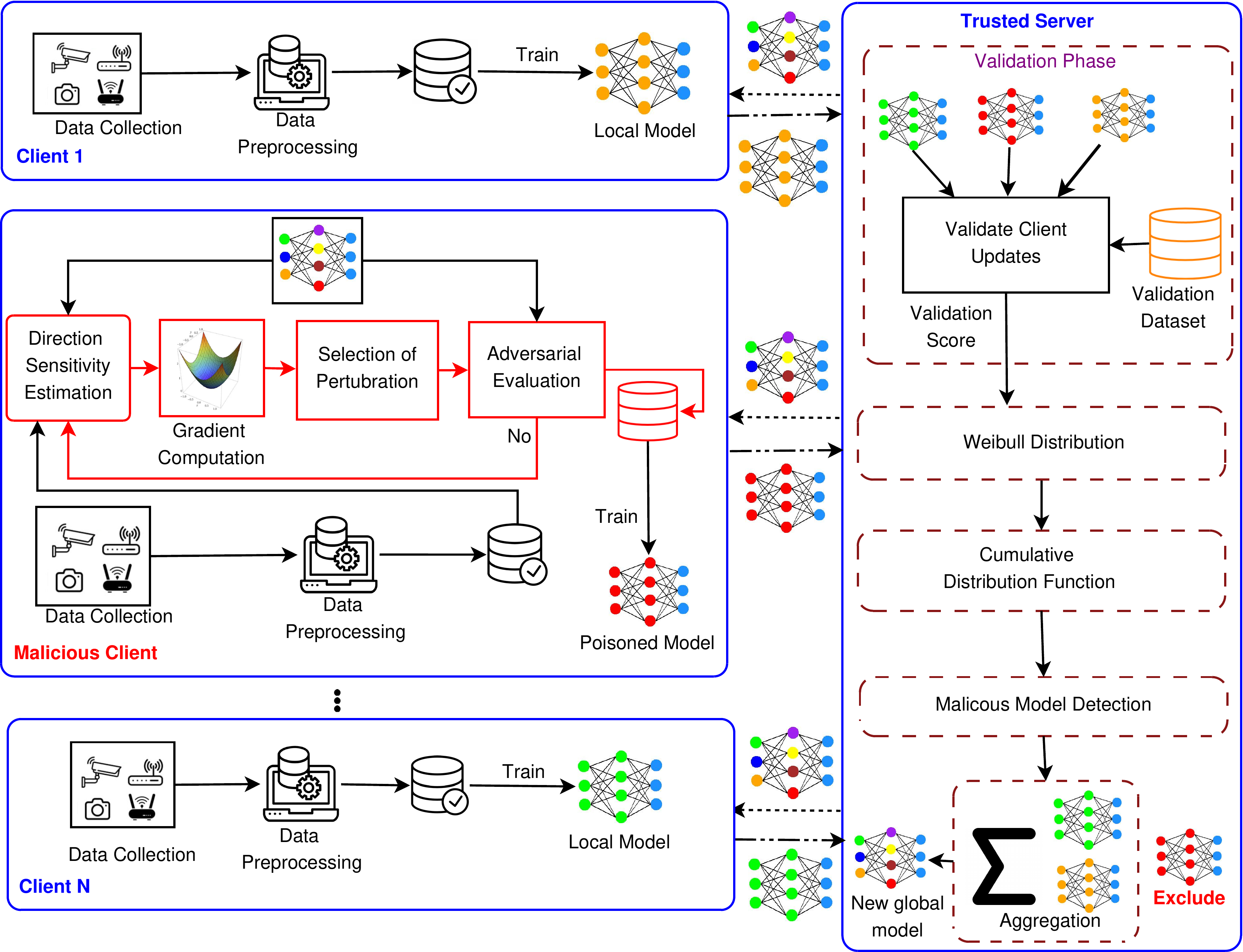}
    \caption{Proposed Architecture}
    \label{fig:weidef_architecture}
\end{figure*}
\subsection{Defense approach for poisoning attack}
\label{subsec:defenseapproach}
FL-based NIDS has emerged as a promising approach to enhance security while preserving data privacy. However, few research efforts have focused on defending against poisoning attacks in this context.
Robust aggregation methods enhance the security of FL-based NIDS by providing more reliable model aggregation than simple averaging, using FedAvg \cite{mcmahan17a}. The authors in \cite{blanchard2017machine} proposed techniques like Krum and Multi-Krum for identifying models closest to the majority using the Euclidean distance method. The coordinate-wise median uses the median of model parameters across all clients, and the trimmed mean computes the global model using the dimension-wise trimmed mean of local models \cite{yin2018byzantine}. Nguyen et al. proposed FLAME, a method that uses model clustering and weight clipping to maintain the benign performance of the aggregated model while effectively eliminating the influence of adversarial participants \cite{nguyen2022flame}.

Thein et al. proposed a defense mechanism, pFL-IDS, that personalizes local models in federated learning-based IDS for heterogeneous IoT data, using a mini-batch logit adjustment loss during local training \cite{THEIN2024182}. This loss helps to adapt the models to the specific data distribution of each client. To detect malicious clients, the system calculates the cosine similarity between local models and the centroid of non-poisoned clients derived by comparing the pre-computed global model and the local models. The authors in \cite{YANG2023103381} introduce a lightweight detection mechanism that identifies anomalous participants using a scoring system based on local model loss and dataset size. The Manhattan distance and clustering algorithm are employed to detect anomalies, resulting in an accuracy of 97.1\% on the CIC-IDS-2017 dataset.

SecFedNIDS combines model-level and data-level approaches for defense mechanisms \cite{ZHANG2022154}. They introduce a gradient-based method for selecting important model parameters for the model-level defense, ensuring effective low-dimensional representations of the uploaded local model parameters. Another study also proposed a two-phase defense DPA-FL mechanism to defend against poisoning attacks~\cite{LAI2023103205}. The first phase eliminates potential attackers by comparing the weight differences between participants. The second phase tests accuracy on a small dataset to identify attackers when the model's accuracy drops. 

\section{ System Model and Threat Model}
\label{sec:systemmodel}
\subsection{System Model}
This section presents an overview of the proposed WeiDetect and addresses the inherent poisoning attacks in FL-based network intrusion detection systems. Additionally, we incorporated a robust mechanism to detect and mitigate malicious participants in a heterogeneous computing environment. Figure \ref{fig:weidef_architecture} illustrates the architectural framework of WeiDetect. In this framework, local clients serve as security gateways, capturing IoT network traffic to develop specialized intrusion detection models.

\par The proposed framework comprises a Trusted Server ($TS$) and $N$ collaborative clients \{$\mathscr{C}=\mathscr{C}_1,\mathscr{C}_2 \dots \mathscr{C}_N$\}. Each client $\mathscr{C}_i$ maintains a private dataset $D_i$ with $|D_i|$ samples. Each $D_i$ consists of pairs $(x_{k}, y_{k})$ for indices $k = 1, 2, \ldots, |D_i|$. Moreover, for any two datasets $D_i$ and $D_j$ (different sizes) corresponding to clients $\mathscr{C}_i$ and $\mathscr{C}_j$, are distinct in terms of their data samples. These clients $\mathscr{C}_i$ collaboratively train their local models (named $L_i$) to build effective intrusion detection systems in heterogeneous data distributions. The trusted server $TS$ aggregates these local models to create a global learning model $G$.

\subsection{Threat Model}
In FL-based NIDS, ensuring the reliability of distributed participants is a critical security concern. In collaborative training of FL-based NIDS within IoT environments, malicious clients may compromise edge servers by crafting data poisoning attacks. This can affect the global model's integrity and performance~\cite{nguyen2022flame}.

\subsubsection{Attackers' goal}
In this system, the adversary's primary objective is to compromise the global intrusion detection model by making it produce incorrect predictions. Our research investigates an attack model targeting FL-based NIDS through poisoning attacks that aim to alter the original data distribution. Specifically, adversaries inject malicious samples for a particular class (targeted class) of the compromised local dataset, as detailed in Section \ref{sec:poisoningattack}. This deliberate manipulation increases misclassification rates for the targeted class, making attack detection more challenging. For instance, the intrusion detection system might misclassify malicious traffic as benign, compromising overall security.
\subsubsection{Attackers' capability}
This study considers an adversary operating as an insider. The adversary can access and modify the traffic data in the training datasets of compromised participants to achieve their objectives but cannot interfere with the local model training process or alter the trained model parameters. Notably, any poisoning of traffic data must adhere to traffic domain constraints to maintain its functionality and integrity \cite{ZHANG2022154,hashemi2019towards}. Additionally, the adversary can observe the local model updates of compromised participants and monitor global model changes during each training round. However, the adversary cannot access the server's model initialization or aggregation process. 

Furthermore, the adversary cannot infer the training processes or access the training datasets of other benign clients. This research emphasizes a scenario where the number of compromised participants is limited to a maximum of one-third of the total participants. Studies in \cite{cao2019understanding, blanchard2017machine} indicate that the federated averaging algorithm becomes ineffective beyond this threshold, and the algorithm will not converge. 
To mitigate this risk, we enforce the Byzantine rule, ensuring at least two-thirds of the participants are honest \cite{miao2022privacy}.

\subsubsection{Attackers' strategy}
The attackers employ data injection and label flipping to poison the samples. In both cases, the attacker manipulates samples from the targeted class within the training dataset. In data injection, the targeted class samples are crafted using the Fast Gradient Sign Method (FGSM), Projected Gradient Descent (PGD), and Gaussian Noise (GN) before local training. 
In label flipping, samples from the targeted class are flipped to another class, and to evade detection, the attacker will focus on multiple targeted classes.
\subsection{Poisoning attacks in FL-based NIDS}
\label{sec:poisoningattack}
FL-based NIDS exemplifies the advantage of leveraging distributed network security while preserving data privacy. However, this distributed framework also introduces vulnerabilities, as adversaries can exploit the collaborative learning process. Specifically, the intrusion detection model, updated iteratively with incoming network traffic data, is susceptible to data poisoning attacks. In this context, a malicious client ($C_{i}$), controlled by an adversary, introduces a manipulated dataset $\tilde{D}_i$ to contaminate the global model. The attack's impact through this strategic model contamination during training is given in Eq.~\ref{eq:poisonlocalupdates}.
\begin{equation}
\label{eq:poisonlocalupdates}
    \tilde{L}_i^{(t)} = L_i^{(t)} - \alpha \nabla  \mathcal{L}_i(L_i^{(t)}, \tilde{D}_i)
\end{equation}
Where $\tilde{L}_i^{(t)}$ represents the contaminated model weight trained on a compromised dataset, the malicious participant transmits these model parameters to the server $TS$. 

This study investigates a sophisticated adversarial strategy targeting the FL systems through class-specific data poisoning. To increase the stealthiness of the attack, the adversary modifies a proportion \( P \) of the total training samples for poisoning, targeting a specific class to evade detection mechanisms. The primary goal of such targeted attacks is to degrade the global model's performance in a specific class while maintaining accuracy in other cases, thereby reducing the likelihood of detection.

In the local training phase, the $C_i$ generates targeted adversarial samples \( x_k^{adv} \) for the sample \( x_k \) belonging to a specified target class $TL$. For a given input \( x_k \), its corresponding label \( y_k \), and the model parameters $\theta_i$ of client $C_i$ ,the adversarial perturbation \( \delta_k \) is computed to maximize the loss function:

\begin{equation}
\delta_k = \arg\max_{\delta} \ell(L_i(x_k + \delta;\theta_i), y_k),
\end{equation}
where \( \ell(\cdot) \) denotes the loss function, and \( L_i(x_k + \delta;\theta_i) \) represents the output of the model $L_i$ for the client $C_i$. The generated adversarial sample \( x_k^{adv} \) is given by:

\begin{equation}
x_k^{adv} = x_k + \delta_k.
\end{equation}

We explore three widely adopted adversarial sample generation methods, the Fast Gradient Sign Method (FGSM), Projected Gradient Descent (PGD), and Gaussian Noise (GN), to study the impact on the system. Each method introduces adversarial perturbations \( \delta_k \) in unique ways, as described below.
\begin{itemize}
    \item  \textit{Fast Gradient Signed Method} \cite{goodfellow2014explaining} is a single-step attack that computes the adversarial perturbation \( \delta_k \) by maximizing the loss function with respect to the input \( x_k \). The perturbation is calculated as:

    \begin{equation}
    \delta_k = \epsilon \cdot \text{sign} \left( \nabla_{x_k} \ell(L_i(x_k; \theta_i), y_k) \right)
    \end{equation}
    
    where \( \epsilon \) controls the magnitude of the perturbation, \( \text{sign}(\cdot) \) denotes the sign function, and \( \nabla_{x_k} \ell(\cdot) \) represents the gradient of the loss function concerning \( x_k \). 
    \item  \textit{Projected Gradient Descent} \cite{madry2017towards} is an iterative extension of FGSM that repeatedly adds perturbations to the input \( x_k \) while ensuring that the perturbations remain within an allowable bound. The perturbation at each iteration \( r \in R \) is computed as:
    \begin{equation}
    x_k^{r+1} = \Pi_{x_k + S} \left( x_k^r + \alpha \cdot \text{sign} \left( \nabla_{x_k^r} \ell(L_i(x_k^r; \theta_i), y_k) \right) \right)
    \end{equation}
    
    where \( \alpha \) is the step size, \( \Pi_{x_k + S}(\cdot) \) is the projection function to ensure that \( x_k^{r+1} \) remains within the perturbation budget \( S \), and \( x_k^0 = x_k \). The final adversarial sample $(x_k^{adv})$ is obtained after a predefined number of iterations ($R$).
    \item \textit{Gaussian Noise} \cite{triastcyn2019federated} introduces randomness into the input by adding noise from a Gaussian distribution. The perturbation is defined as:

    \begin{equation}
    \delta_k \sim \mathcal{N}(\mu, \sigma^2),
    \end{equation}
    
    where \( \mu \) and \( \sigma^2 \) are the mean and variance of the Gaussian data distribution. Unlike FGSM and PGD, GN does not rely on the gradient of the loss function.
    
\end{itemize}

We chose these adversarial methods due to their presence in existing state-of-the-art works \cite{ZHANG2022154,yamany2021oqfl}, highlighting the resilience of the proposed framework against such attacks. Our study targets data poisoning attacks, given their ability to severely undermine model integrity. 

Once the adversarial samples are generated and tested against the current global model \( G^t \), if the sample \( x_i^{adv} \) is misclassified, it is added to the adversary’s local dataset. 
Eventually, the attackers introduce the poisoned traffic data into the local training dataset to create a poisoned dataset \( \tilde{D}_i\) and then train the model to produce a compromised intrusion detection model ($\tilde{L}_i$). These corrupted models ($\tilde{L}_i$) are subsequently uploaded to the trusted server to subvert the global intrusion detection model.
\begin{figure}[h!]
    \centering
    \includegraphics[width=0.9\linewidth]{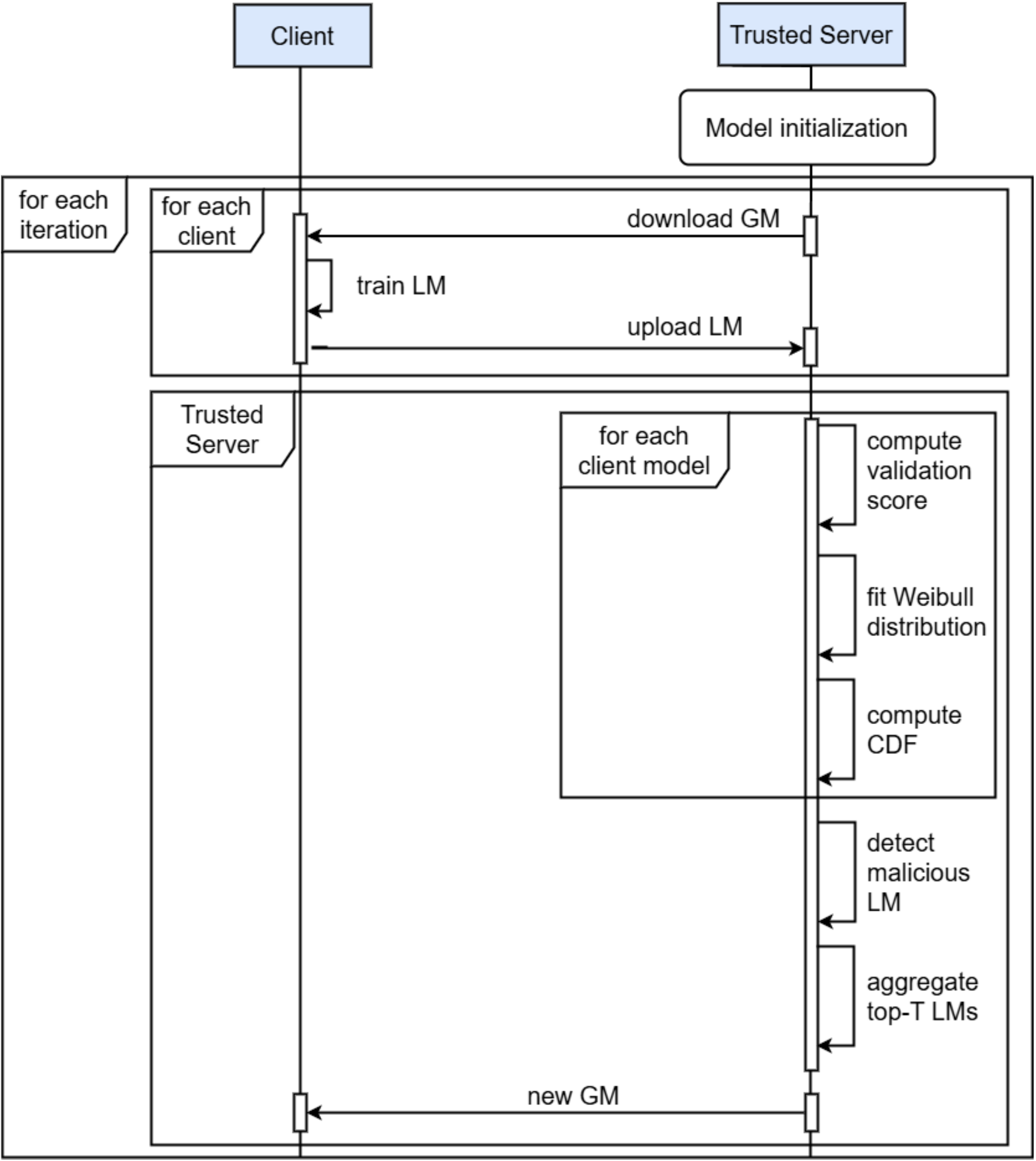}
    \caption{WeiDetect sequence diagram.}
    \label{fig:weidef_sequence_diagarm}
\end{figure}

\section{WeiDetect: Proposed defense mechanism}
\label{sec:proposedmethod}
\subsection{Overview}
This section presents an overview of the proposed FL-based approach for NIDS. The adversary aims to degrade the performance of the intrusion detection model through data poisoning attacks, triggering the model to make unpredictable results for specific tasks. To counter the malevolent impact on the global model, we introduce a novel defense mechanism, WeiDetect, designed to work with non-IID data distribution scenarios in FL-based NIDS. 

WeiDetect operates on the server side, detecting and removing malicious or low-quality models before the global model aggregation. WeiDetect employs a two-phase defense strategy. The first phase, \texttt{validation phase}, evaluates the validation score of the local models, as illustrated in section \ref{sec:validationscore}. The second phase, \texttt{detection phase}, estimates anomalous models using the Weibull probability distribution, as depicted in \ref{sec:anomallyweibull}. 

\sloppypar{Figure \ref{fig:weidef_sequence_diagarm} shows the sequence diagram of the proposed WeiDetect. In each round, clients download the Global Model (GM), train locally, and upload their updated Local Model (LM). The Trusted Server ($TS$) evaluates each LM on an auxiliary dataset (only access to the $TS$). Subsequently, the $TS$ fits validation scores to a Weibull distribution and extracts shape and scale parameters. It then computes the CDF and selects the top $\mathcal{T}$ LMs based on CDF, and aggregates them to generate a new GM. The updated GM is then distributed to clients for the next round, ensuring continuous improvement. }

\begin{algorithm}[h!]
\caption{Proposed WeiDetect algorithm}
\label{alg:WeiFighter_defense}
\begin{algorithmic}[1]
\renewcommand{\algorithmicrequire}{\textbf{Input:}}

\Require Total number of clients $N$; Number of aggregation clients $M$; Communication rounds $T$; Server auxiliary dataset $AD$; Adversary controlled clients $\mathcal{A}$; Clients local datasets $\{D_1, D_2, \dots, D_N\}$; Target Label $TL$; 
\renewcommand{\algorithmicrequire}{\textbf{Output:}}
\Require  Global model $G^{T}$

\Statex \textbf{Server executes:}
\State $G^0 \leftarrow$ initial global model
\For {each communication round $t = 1, 2, \dots, T$}
    
    \State $SClients \leftarrow$ Randomly selected set of $M$ clients
    \For{each client $k \in SClients$}
        
        \State $\hat{L}_k^{t} \gets ClientUpdate(k, G^{t-1})$ \Comment{Local updates}
    \EndFor
   
    \State $BenignClients \gets WeibullFiltering(\hat{L}_k^{t} | k \in SClients)$
   
     \Statex \Comment{ Malicious or low-quality model detection and extract benign client}
    \State $G^{t} \gets$ aggregates models in   $BenignClients$  using Eq.~(\ref{eq:fedavg}) \Comment{Aggregation phase}
\EndFor



        


\end{algorithmic}
\end{algorithm}

 Due to poor data quality or insufficient samples for specific classes (data imbalances) in heterogeneous environments, legitimate models become low-quality, similar to malicious models. While these low-quality models are not malicious by design, they still degrade the global model's performance. Since our goal is to exclude the poisoned or low-quality model from aggregation, it is essential to identify them and minimize their impact on the global model. Therefore, validating models \{${\hat L_1^{(t)}, \ldots, \hat L_K^{(t)}}$\} in each federation round $t$ before aggregation can help identify suspicious adversarial participants. The detailed design of our proposed approach is illustrated in Algorithm~\ref{alg:WeiFighter_defense}.

\begin{algorithm}
\caption{ClientUpdate}
\label{alg:ClientUpdate}
\begin{algorithmic}[1]
    \renewcommand{\algorithmicrequire}{\textbf{Input:}}
    \renewcommand{\algorithmicensure}{\textbf{Output:}}
    
    \Require Global model  $G^{t-1}$ 
    \renewcommand{\algorithmicrequire}{\textbf{Output:}}
    \Require client models $\hat L_k^{(t)}$ 
    \State $Adv^t \leftarrow$ Set of Adversaries
\State $D_k \leftarrow$ Local dataset held by $C_k$  


    \If{$k \in Adv^t$} \Comment{Check if client $k$ is an adversary}
        \State $I_T \gets$ Samples from $D_k$ with target label $TL$ 
         \State  $D^{nontarget}_K \leftarrow $ Non-target samples from $D_k$
        
        \State $L'_i \gets$ Train model $G^{t-1}$ using $D^{nontarget}_K$
        \State $I_A \gets$ Generate adversarial samples for $I_T$ 
        \For{each adversarial sample $I_a \in I_A$ }
            \State $\hat{y} \leftarrow \text{Test } I_a \text{ using global model } G^{t-1}$
            \If{$\hat{y}$ is not equal to $TL$}
              \State Replace sample $I_T$ in $D_k$ with $I_a$
            \EndIf
        \EndFor
    \Else
        \State $\hat{L_i} \gets$ Train model $G^{t-1}$ on $D_K$
    \EndIf
\State \Return $\hat{L_i}$
\end{algorithmic}
\end{algorithm}

During federation rounds $t$, the server randomly selects $M$ clients, as specified in line 3. Each selected client trains $G^t$ using their private dataset $D_i$. The compromised participants execute the targeted data poisoning attack through adversarial sample generation to poison the dataset, as outlined in Algorithm~\ref{alg:ClientUpdate} ( lines 3 to 14). Then updates the local model $\hat L_i^{(t)}$ using a stochastic gradient descent optimization algorithm to minimize a loss function $\mathcal{L}_i$. This process is described mathematically in Eq.~(\ref{eq:localupdates}),
\begin{equation}
\label{eq:localupdates}
    \hat L_i^{(t)} = L_i^{(t)} - \alpha \nabla  \mathcal{L}_i(L_i^{(t)}, D_i)
\end{equation}
Here $\alpha$ denotes the learning rate, and $\nabla  \mathcal{L}_i(L_i^{(t)}, D_i)$ is the gradient of local optimization loss computed using the local dataset $ D_i$.
These local model's ($ \hat L_i^{t}$) parameters are transmitted to a central $TS$.

On the server side, instead of indiscriminately aggregating all local models, the WeiDetect defense approach employs a mechanism that detects malicious or low-quality models based on validation scores and the Weibull distribution. As outlined in lines 1 to 9 of Algorithm~\ref{alg:poisoingdetctor}, this method effectively models the tail behavior of reliability scores to identify potential adversarial clients.

These clients are identified as potential adversaries or low-quality participants, as their models exhibit performance characteristics that deviate significantly from the expected distribution. Subsequently, the server removes these models and aggregates the selected local models \{${\hat L_1^{(t)}, \ldots, \hat L_m^{(t)}}$\}, as shown in line 9 of Algorithm~\ref{alg:WeiFighter_defense}. The selected models are then used to synthesize a unified global model $G^{t+1}$, as shown in line 10, using the equation given in Eq.~(\ref{eq:fedavg}).  

The trusted server $TS$ redistributes $G^{t+1}$ to all participating clients to initialize their local models for the next round of training, as shown in Eq.~(\ref{eq:dissimination}).
\begin{equation}
\label{eq:dissimination}
 L_i^{(t+1)} \leftarrow G^{t+1}, i \in \{1,2,\dots, N\}
\end{equation}

\begin{equation}
\label{eq:fedavg}
    G^{t+1}= \frac{1}{m}\sum_1^m \hat L_i^t
\end{equation}

The process is repeated across multiple communication rounds until the global intrusion detection model converges or the predefined number of epochs is reached.
\par The following subsections provide a detailed description of the phases of the WeiDetect defense.

\begin{algorithm}
\caption{WeibullFiltering}
\label{alg:poisoingdetctor}
\begin{algorithmic}[1]
    \renewcommand{\algorithmicrequire}{\textbf{Input:}}
    \renewcommand{\algorithmicensure}{\textbf{Output:}}
    
    \Require Clients models list $\hat{L^{(t)}} \leftarrow {\hat L_1^{(t)}, \ldots, \hat L_K^{(t)}}$\} in round $t$, $AD:$ Auxiliary Dataset
    \renewcommand{\algorithmicrequire}{\textbf{Output:}}
    \Require $\text{min\_client:}$ List of malicious or low-quality clients 
    \State Initialize $ValScore$ as an empty list
    \For{each client model in $\hat{L^t}$} \Comment{Validation Phase}
        \State $ValScore_k \leftarrow$ Test $\hat{L}_k^{t}$ using $AD$ and record the F1-Score 
        
    \EndFor
  
    \State $floc \gets $location parameter for Weibull fitting \Comment{Detection Phase}

    \State $\text{params} \gets \text{Fit Weibull distribution to the} $ $ValScore$

    \State Extract shape and scale parameters from params

    \State $\text{CDFVal} \gets$Compute the CDF for the Weibull distribution using shape and scale parameters

    \State $\text{benign\_client} \gets \text{Select top~}$$ \mathcal{T}$$ \text{client with highest CDF values}$

    \State \textbf{Return} $\text{benign\_client}$
\end{algorithmic}
\end{algorithm}

\subsection{Validation phase}
\label{sec:validationscore}

Adversary aims to undermine an FL model's performance by poisoning compromised participants' datasets, causing their performance deviation. Since FL relies on aggregating local updates, the impact of these malicious updates becomes more pronounced, especially in highly imbalanced non-IID data distribution scenarios. Validating clients local updates (\{${\hat L_1^{(t)}, \ldots, \hat L_K^{(t)}}$\}) before aggregation will assist the truster server for identifying suspicious adversarial clients and mitigate their impact.

We propose the \texttt{Local Validation function}, $\mathcal{V}$, to evaluate the performance of each $\hat L_i^{(t)}$. The $\mathcal{V}$ function uses a fixed validation set (auxiliary dataset $AD$) stored on the server, constructed using prior knowledge or expert insights. To generate $AD$, we first pinpoint instances with high confidence levels by assessing their softmax probabilities. Accurately classified data instances that exceed a predefined probability threshold are selected for $AD$. Specifically, instances with a probability greater than $0.9$ are included in the dataset. The $\mathcal{V}$ function computed for the client model $\hat L_i$ is formally defined as follows.
\begin{equation}
\label{eq:validationscore}
    \mathcal{V}_{i} =\text{F1-score}(\hat L_i, AD)
\end{equation}
where $\text{F1-score}(\hat L_i, AD)$ represents the standard F1-score evaluation metric for the local model $\hat L_i$ on the auxiliary dataset $AD$. The detailed steps of this process are illustrated in lines 2 to 4 in Algorithm~\ref{alg:poisoingdetctor}.

\subsection{Anomaly detection phase using Weibull distribution model}
\label{sec:anomallyweibull}

The non-IID nature of data in most FL settings makes it challenging to identify client types in advance, making it difficult to determine the exact number of adversarial clients. Additionally, the presence of low-quality models cannot be eliminated. Benign client models are expected to converge to a common solution, whereas adversarial or low-quality models tend to deviate, resulting in suboptimal convergence and behaving like outliers. Consequently, relying solely on the validation score is insufficient to differentiate between these types of models. 

The Weibull distribution offers a robust probabilistic framework for outlier detection due to its adaptability in modeling diverse data distributions. Its ability to manage data sparsity and detect anomalous patterns makes it well-suited for high variability and distribution shift scenarios. Prior studies \cite{ndeba2025comparative,yuen1985comparisons,zhang2022parameter}, have demonstrated its effectiveness in outlier detection, while \cite{surucu2009monitoring} highlights its application in monitoring reliability within the manufacturing industry. Additionally, the Weibull distribution has been widely used in various domains, including survival analysis \cite{rahman2023fedpseudo,bennis2021dpwte} and fault prediction \cite{xu2023multi}, further establishing its relevance for identifying deviations and anomalies in complex datasets. 

The authors in \cite{marfo2025adaptive} employed the Weibull distribution for adaptive client selection and checkpointing to improve network anomaly detection performance and fault tolerance. This motivates us to leverage the Weibull distribution model for detecting outlier models in federated learning settings, where non-IID data distributions and adversarial manipulations can lead to significant deviations in model performance. Due to its flexibility in capturing distributional shifts and identifying anomalies, we aim to enhance the system's robustness using Weibull distribution. Furthermore, the probabilistic nature of the Weibull distribution model detects benign and adversarial client models, providing a reliable method for outlier detection in dynamic and heterogeneous environments. 

The following subsections provide a detailed description of the anomaly detection phase using the Weibull distribution model.

\subsubsection{Weibull distribution formulation} 
To detect clients whose models exhibit anomalous behavior, we fit an exponential Weibull distribution to the validation scores of participating models. The exponential Weibull distribution's probability density function (PDF) is calculated using Eq. \ref{eq:pdf}.
\begin{equation}
\label{eq:pdf}
f(\mathcal{V}_i; \lambda, \beta) = \frac{\beta}{\lambda} \left(\frac{\mathcal{V}_i}{\lambda}\right)^{\beta-1} e^{-\left(\frac{\mathcal{V}_i}{\lambda}\right)^\beta}, \quad \mathcal{V}_i > 0
\end{equation}

where \( \mathcal{V}_i \) represents the validation score (F1-score) of the local model \( \hat L_i \). The parameter \( \beta\) is the shape parameter that controls the form of the Weibull distribution, while \( \lambda \) is the scale parameter that determines the range of values. The term \( \left(\frac{\mathcal{V}_i}{\lambda}\right)^{\beta-1} \) ensures that the function scales appropriately according to the shape parameter. The exponential term \( e^{-\left(\frac{\mathcal{V}_i}{\lambda}\right)^\beta} \) defines the PDF, ensuring that it follows the Weibull distribution. Finally, the condition \( \mathcal{V}_i > 0 \) ensures that the function is valid only for positive validation scores.

\subsubsection{Parameter estimation} 
Using maximum likelihood estimation, the model fits the distribution to the validation accuracy scores, extracting the parameters $\hat{\lambda}, \hat{\beta}$ are as follows:

\begin{equation}
    \hat{\lambda}, \hat{\beta} = \arg\max_{\lambda,\beta} \prod_{i=1}^{N} f(\mathcal{V}_i; \lambda, \beta)
\end{equation}
Where, \( N \) represents the total number of clients in the aggregation. \( \mathcal{V}_{i} \) is the validation score of the \( i \)-th client. \( \alpha, c, \lambda, \theta \) are the parameters of the exponentiated Weibull distribution.

\subsubsection{Computing Cumulative Distribution Function (CDF)}
The CDF of the exponential Weibull distribution is used to measure how extreme each client's validation score is:
\begin{equation}
    F(\mathcal{V}_{i},\hat{\lambda}, \hat{\beta}) = \left( 1 - exp^{-\left( \frac{\mathcal{V}_{i}}{\lambda}\right)^\beta} \right).
\end{equation}
This function provides the probability of a randomly chosen client's validation score.

\subsubsection{Anomalous model detection}  
The WeiDetect framework improves the robustness of federated learning by systematically filtering out low-quality or adversarial models. To achieve this, it selects the $\mathcal{T}$ clients with the highest cumulative distribution function (CDF) values for aggregation, ensuring that only reliable and well-aligned models contribute to the global model. The selection of clients with the highest CDF values is motivated by their stronger alignment with the expected benign data distribution (Eq.~\ref{eq:cdf}). A higher CDF value indicates that a client exhibits behavior consistent with trustworthy participants, making its model updates more reliable. In contrast, malicious or low-quality clients display irregular patterns, often resulting in lower CDF values. By prioritizing clients with higher CDF values, WeiDetect mitigates the risk of incorporating adversarial updates and enhances the stability and resilience of the FL-based NIDS system.

\begin{equation}
\label{eq:cdf}
    M^{*} = \arg\max_{M^{*} \subseteq M} \sum_{i \in M} F(\mathcal{V}_{i},\hat{\lambda}, \hat{\beta})    
\end{equation}
\begin{equation*}
    \text{subject to } |M^{*}| = \mathcal{T}
\end{equation*}

where \( \mathcal{M} \) represents the set of all clients in the agagregation, \( M^* \) is the subset of the top \(\tau\) selected clients from \(M \) for aggregation. Algorithm~\ref{alg:poisoingdetctor} lines 5 to 9 depicted the anomaly detection using the Weibull model in WeiDetect.  


\section{Experiment Evaluation and Analysis}
\label{sec:experimentalsection}
This section details the datasets, experimental setup, and evaluation metrics used in our study. We then evaluate the WeiDetect's performance, comparing it with state-of-the-art methods. For experimentation, we used the PyTorch framework on a system equipped with an Intel Core i9 processor, 5GB NVIDIA GPU, and 32GB of RAM. The adversarial attack was launched using the Torchattacks library in PyTorch to generate adversarial samples.

\subsection{Datasets and model architecture}
In this study, we utilized two benchmark network traffic datasets to evaluate the performance of the proposed WeiDetect framework. We selected a subset of datasets used in prior FL-based NIDS \cite{nguyen2020poisoning,TABASSUM2022299,LAI2023103205} for evaluation.
\begin{itemize}
    \item \textbf{CIC-Darknet2020:} The CIC-Darknet2020 dataset is a real-world network traffic dataset created by merging two public datasets from the University of New Brunswick: ISCXTor2016 and ISCXVPN2016 \cite{rust2023darknet}. Each sample in CIC-Darknet2020 represents traffic features derived from raw packet capture sessions. The dataset includes 141,530 hierarchically labeled samples categorized into four top-level traffic types: Non-Tor (93,356 samples), Tor (1,392 samples), VPN (22,919 samples), and Non-VPN (23,863 samples). We preprocessed the dataset to eliminate irrelevant or redundant data, remove missing values, and apply one-hot encoding for categorical features.
    \item \textbf{CSE-CIC-IDS2018:} The CSE-CIC-IDS2018 dataset includes seven attack scenarios such as Heartbleed, DDoS, Botnet, Infiltration, Web, DoS, and Brute-force attacks, capturing network traffic and system logs with 83 features extracted via CICFlowMeter-V3. The dataset is labeled with different attack types, such as DoS Golden Eye, Heartbleed, DoS Hulk, DoS Slow HTTP, DoS Slowloris, DDoS, SSH-Patator, FP, Patator, Brute Force, XSS, Botnet, Infiltration, PortScan, and SQL Injection. Further, we removed features like SrcPort, Flow ID, Timestamp, IP addresses, and duplicate rows. Finally, we used a dataset consisting of 575,364 DDoS-LOIC-HTTP, 198,861 DDoS-HOIC, 145,199 DoS-Hulk, 144,535 Bot, 140,610 Infiltration, 94,048 SSH-Bruteforce, 41,406 DoS-GoldenEye, and 9,908 DoS-Slowloris samples. Additionally, we performed preprocessing on the dataset by removing missing values, applying one-hot encoding for categorical features, and normalizing numerical data to eliminate irrelevant or redundant information.
\end{itemize}
To assess the performance, we split each dataset into 80\% for training and 20\% for testing. We further partition the training portion and distribute it among clients to train local intrusion detection models based on the Dirichlet parameter.

\begin{table}[h!]
\centering
\caption{Hyperparameters and model architecture setting for each dataset.}
\label{tab:empty_model_settings}
\renewcommand{\arraystretch}{1.2}
\resizebox{0.5\textwidth}{!}{ 
\begin{tabular}{lp{2.55cm}p{2.8cm}}
\hline
Dataset $\rightarrow$& CIC-Darkent2020 & CSE-CIC-IDS2018\\ \hline
\multicolumn{3}{l}{Federated learning parameters} \\ \hline
Number of client&  20  &  20    \\ 
Local epochs&   5 &    3  \\
FL round&   100 &    50  \\ \hline
\multicolumn{3}{l}{Model parameters} \\ \hline
Layers&HL:(64,32); OL:4&HL:(64,32); OL:8 \\
Activation function& ReLu&ReLu \\
Optimizer& SGD&SGD \\
Learning rate & 0.01&0.001 \\
 Batch size & 32&64 \\
 \hline
\end{tabular}}
\caption*{\footnotesize{HL denotes the hidden layer and OL represents the output layer}}
\end{table}
\begin{figure}[h!]
\centering
\begin{subfigure}{0.45\textwidth}
    \includegraphics[width=0.9\linewidth]{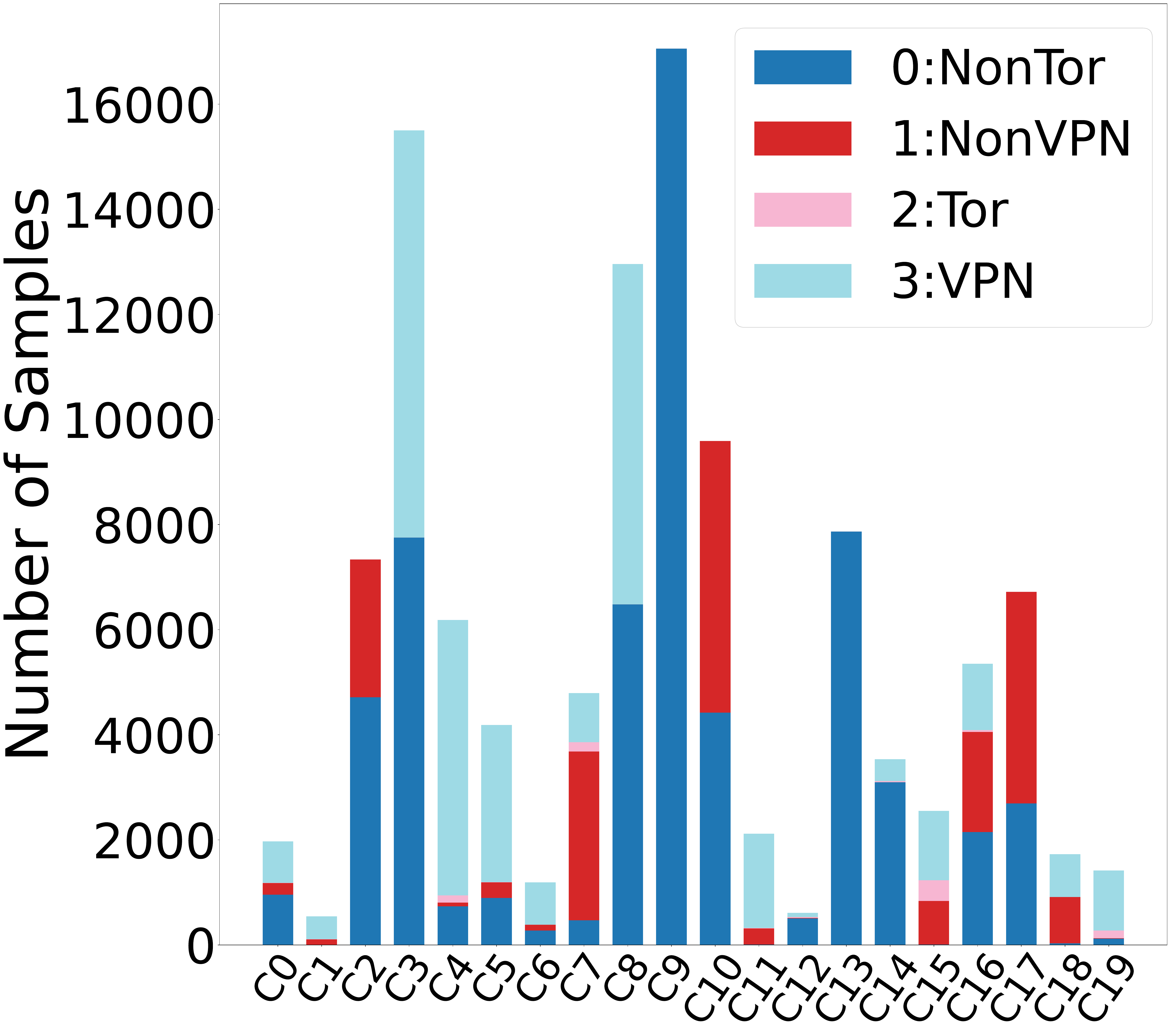} 
    \caption{CIC-Darknet2020}
    \label{fig:cicdarknet_noniid}
\end{subfigure}
\hfill
\begin{subfigure}{0.45\textwidth}
    \includegraphics[width=0.9\linewidth]{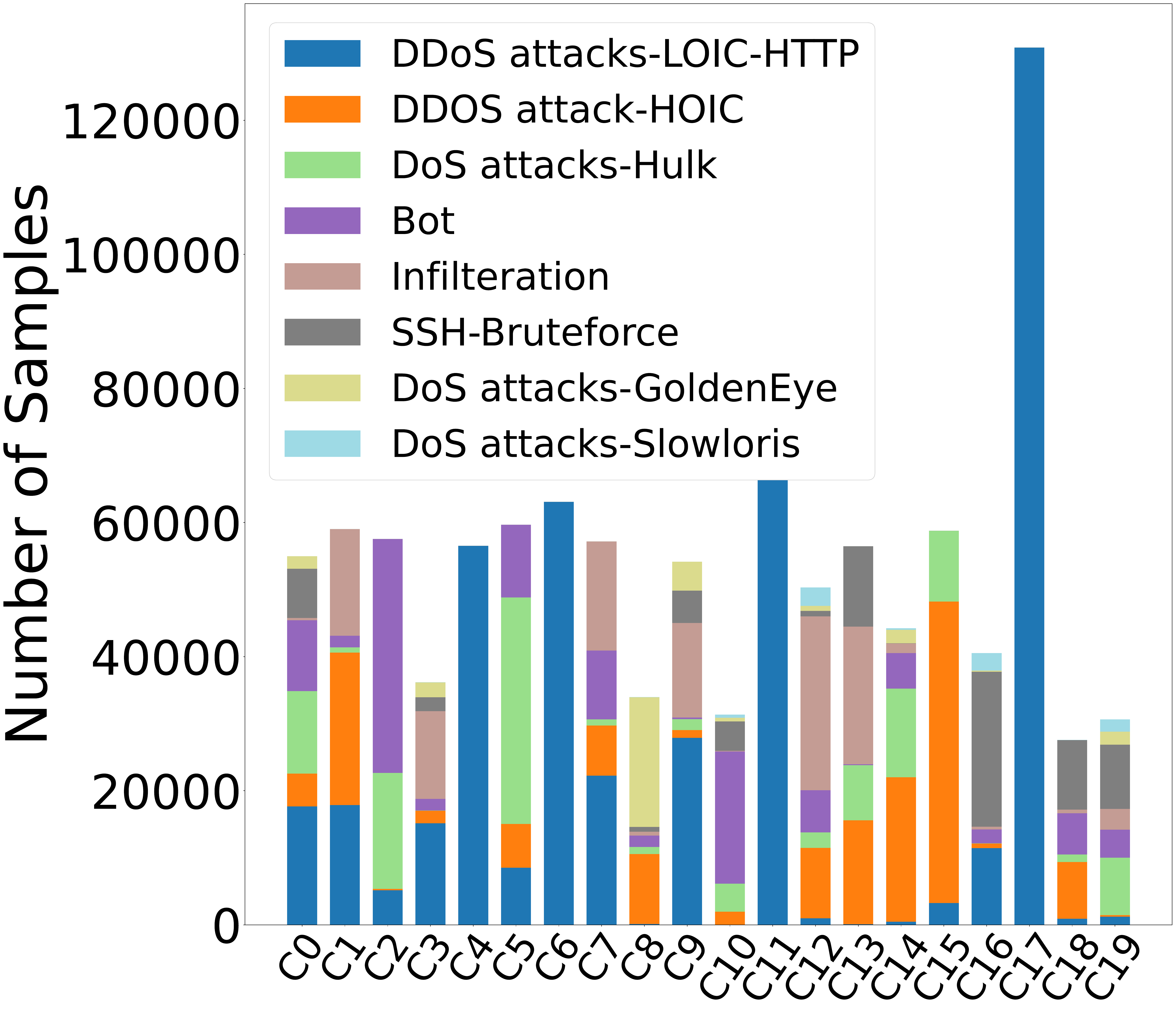} 
    \caption{CIC-IDS2018}
    \label{fig:cicids2018_noniid}
\end{subfigure}
\hfill
\caption{A schematic illustration of the samples per class allocated to each client in different datasets. The x-axis represents the client IDs, and the y-axis represents the number of samples.}
\label{fig:datadistribution}
\end{figure}

\subsection{FL simulation setup}
For the FL setup, we follow the experimental configuration in \cite{xiong2021privacy} for selecting the number of clients, setting $N = 20$ for the training process.  
A horizontal FL framework with non-IID data distribution, simulated using a Dirichlet distribution with $\eta = 0.4$  is used to create highly skewed data distributions \cite{bagdasaryan2020backdoor}. Previous studies \cite{agrawal2022federated} frequently utilized the Dirichlet distribution to simulate non-IID data distributions in FL. Figure \ref{fig:datadistribution} presents the schematic illustration of per-class data distribution among the clients in different datasets. Table \ref{tab:empty_model_settings} lists the hyperparameters and model architecture for training each dataset.

\par We evaluate the effectiveness of WeiDetect in defending against poisoning attacks under diverse adversarial conditions. In our experimental setup, we set the percentage of malicious participants to 15\% 
to maintain a realistic attack scenario \cite{}. These malicious participants generate adversarial samples for specific classes using established adversarial sample generation techniques such as FGSM, PGD, and GN. We investigate two distinct attack scenarios: a focused single-class poisoning attack and a more comprehensive triple-class poisoning attack. Furthermore, we analyze the impact of varying data poisoning ratios $(10\%, 20\%,30\%,40\%, and 50\%)$ of training data compromised by malicious participants for each targeted class.

In order to generate adversarial samples, we apply FGSM, PGD, and GN approaches. We initially set the perturbation value to $\epsilon=0.35$ for both FGSM and PGD in our experimental setup. However, we also experimented with smaller $\epsilon$ values to ensure that the adversarial samples remained visually closer to the clean samples while maintaining their effectiveness in the attack. 
As $\epsilon$ increases, the perturbations become more pronounced, making the adversarial example more distinguishable from the clean data. The PGD attack generates perturbed samples that retain high visual similarity to the clean sample, whereas FGSM introduces more noticeable deviations. We maintained the default parameters for the GN method's attack settings.
\subsection{Performance metrics}
We evaluated the system's performance using the following metrics to study the impact of the poisoning attack. 
\begin{itemize}

    \item \textbf{F1-score}: We employed the F1-score to assess the FL-based NIDS global model’s performance due to the imbalanced nature of our datasets, which is computed by Eq.~(\ref{eq:f1}).
    \begin{equation}
        \label{eq:f1}
        \text{F1-score}=\frac{TP}{TP+ \frac{1}{2}(FP+FN)}
    \end{equation}
    \noindent $TP$ represents the count of attack samples correctly identified as attacks, and $FP$ denotes the count of benign samples incorrectly identified as attacks. $FN$ exhibits the count of attack samples incorrectly identified as benign.
    \item \textbf{Target Class Accuracy (TAC)}: It quantifies the accuracy of the targeted class ($C_T$) in which the model classifies instances correctly. In this study, the target class refers to the class in which the attackers inject the modified traffic data (poisoning attack). The poisoning attack launched in our experiments is a targeted attack that aims to degrade the performance of the target class and not affect other classes. Eq.~(\ref{eq:tca}) illustrates the mathematical expression for TAC, where $|N_{C_{T}}|$ represents the total number of samples for the class $C_{T}$. 

    \begin{equation}
        \label{eq:tca}
        \text{TCR} = \frac{1}{|N_{C_{T}}|} \sum_{i=1}^{|N_{C_{T}}|} [ (\hat{y}_i = C_{T} \, | \, y_i = C_{T})]
    \end{equation}

\end{itemize}
\subsection{Experimental results}
\label{sec:experimentalresult}
In this section, we evaluated the effectiveness of our proposed WeiDetect against the poisoning attacks on imbalanced data. We first present the results for the baseline scenario under non-adversarial conditions. Then, it analyzed the attack's impact on FL-based NIDS without any defense in the adversarial scenario using the FedAvg approach. Further, we compare the WeiDetect defense results with existing state-of-the-art defense approaches, including Krum, MKrum, Median, Trimmed Mean, and FLAME, under an adversarial environment.
\begin{table}[h!]
    \centering 
    \renewcommand{\arraystretch}{1.3}
    \begin{tabular}{lcc}
        \hline
        Dataset & Accuracy &F1-score
        \\
        \hline
         CIC-Darknet2020 & 90.182\% &89.791 \%  \\
      
        CSE-CIC-IDS2018  &94.805\%&94.184\%\\
         \hline   
    \end{tabular}
    \caption{Performance of the FL-based NIDS system under non-poisoning configuration.}
    \label{tab:baseline_results}

\end{table}

\subsubsection{Baseline}
Table \ref{tab:baseline_results} presents the FL-based NIDS's performance without adversarial interference. In the baseline scenarios, experimental results demonstrate that the global model achieves an accuracy of $90.182\%$ and an F1-score of $89.791\%$ on the CIC-Darknet2020 dataset. When evaluated on the CSE-CIC-IDS2018 dataset, the model exhibits performance with an accuracy of $94.805\%$ and achieved an F1-score of $94.184\%$. 

\begin{table*}[ht]
\centering
\caption{Performance comparison of WeiDetect, evaluated using the F1-score matrix, with SOTA methods under a single targeted class adversarial attack on the CIC-Darknet2020 dataset. The highest scores are highlighted in bold.}
\label{tab:sl_cicdarknet}
\footnotesize
\resizebox{\textwidth}{!}{
\begin{tabular}{ccccccccc}
\toprule

Attack strategy& Poison ratio$\downarrow$& FedAvg* & Krum & MKrum & Median  & TMean& FLAME & WeiDetect \\
\toprule

\multirow{5}{*}{FGSM} &10\%  &75.965&75.525 & 83.422 &  74.775  &77.451& 84.856 &  \textbf{87.014} \\ 

&20\% &77.020& 75.452& 76.216 &74.353 &77.191& 84.630 &  \textbf{86.818}  \\

&30\% &75.569&75.358&76.170&74.702&77.020&84.624&
\textbf{86.659} \\
&40\%&75.987&75.319&75.842&73.413&76.575&83.921&\textbf{87.922}\\
&50\% &74.215&75.343&75.378&73.272&76.482&
83.555&\textbf{88.108} \\
 \hline

\multirow{5}{*}{PGD}&10\%&75.383&75.430&75.792&74.164&77.527&81.384&\textbf{87.688}
\\ 

&20\% &75.318&75.408&75.795&73.798&77.324&79.967&\textbf{86.550}\\

&30\% &75.286&75.365&75.631&73.969&77.348&81.045&\textbf{86.472}
\\
&40\% &75.199&75.318&75.627&73.216&77.080&84.172&\textbf{86.959}
\\ 
&50\% &75.161&75.295&75.697&73.217&77.102&84.134&\textbf{86.546}
\\ \hline 

\multirow{5}{*}{GN}&0\% & 75.274&75.362&75.925&72.850&76.982&85.512&\textbf{86.641}
\\ 
&20\% & 75.195&75.288&75.821&72.057&76.424&85.597&\textbf{85.946}
\\
&30\% & 75.093&75.256&75.631&71.953&76.393&85.0&\textbf{85.011} \\
&40\% & 74.981&75.158&75.815&71.443&75.789&84.851&\textbf{85.0}
\\ 
&50\% & 74.911&75.117&75.700&71.314&76.056&84.674&\textbf{85.706}
\\ 
\bottomrule
\end{tabular}}
\end{table*}

\begin{figure*}[h!]
\centering
\begin{subfigure}{0.33\textwidth}
    \includegraphics[width=\linewidth]{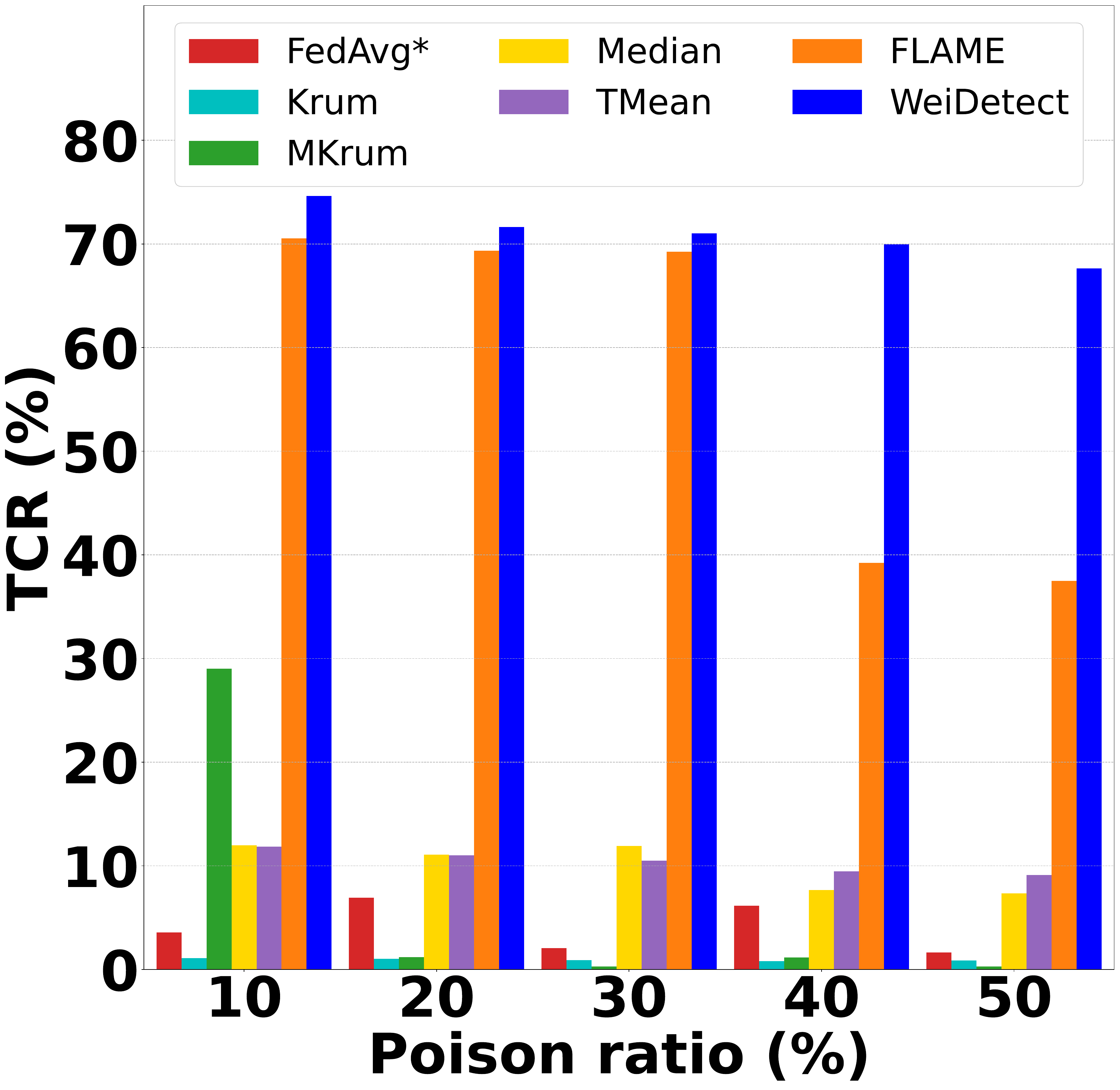} 
    \caption{FGSM}
    \label{fig:cicdarknet_sl_fgsm}
\end{subfigure}
\hfill
\begin{subfigure}{0.33\textwidth}
    \includegraphics[width=\linewidth]{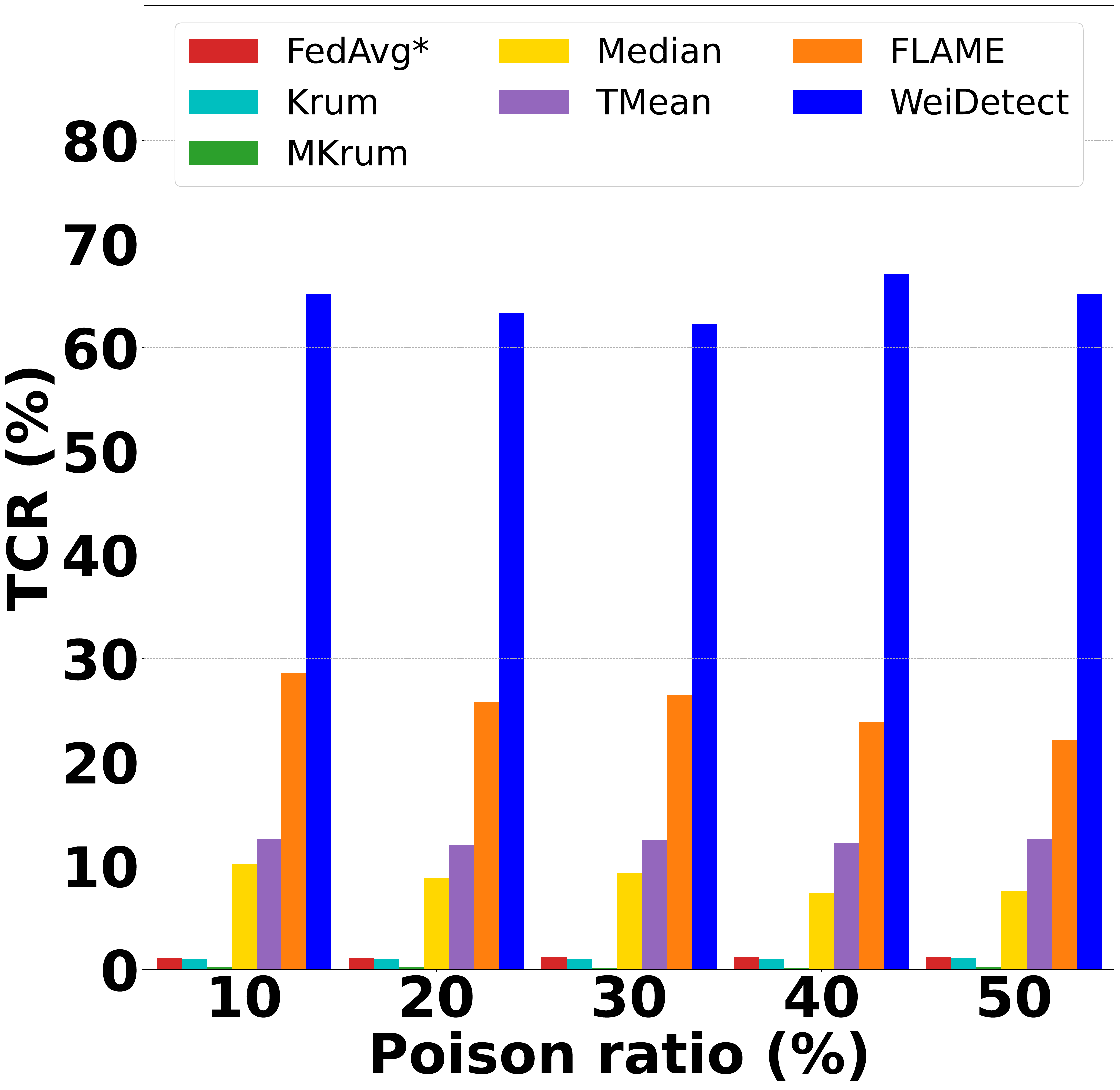} 
    \caption{PGD}
    \label{fig:cicdarknet_sl_pgd}
\end{subfigure}
\hfill
\begin{subfigure}{0.33\textwidth}
    \includegraphics[width=\linewidth]{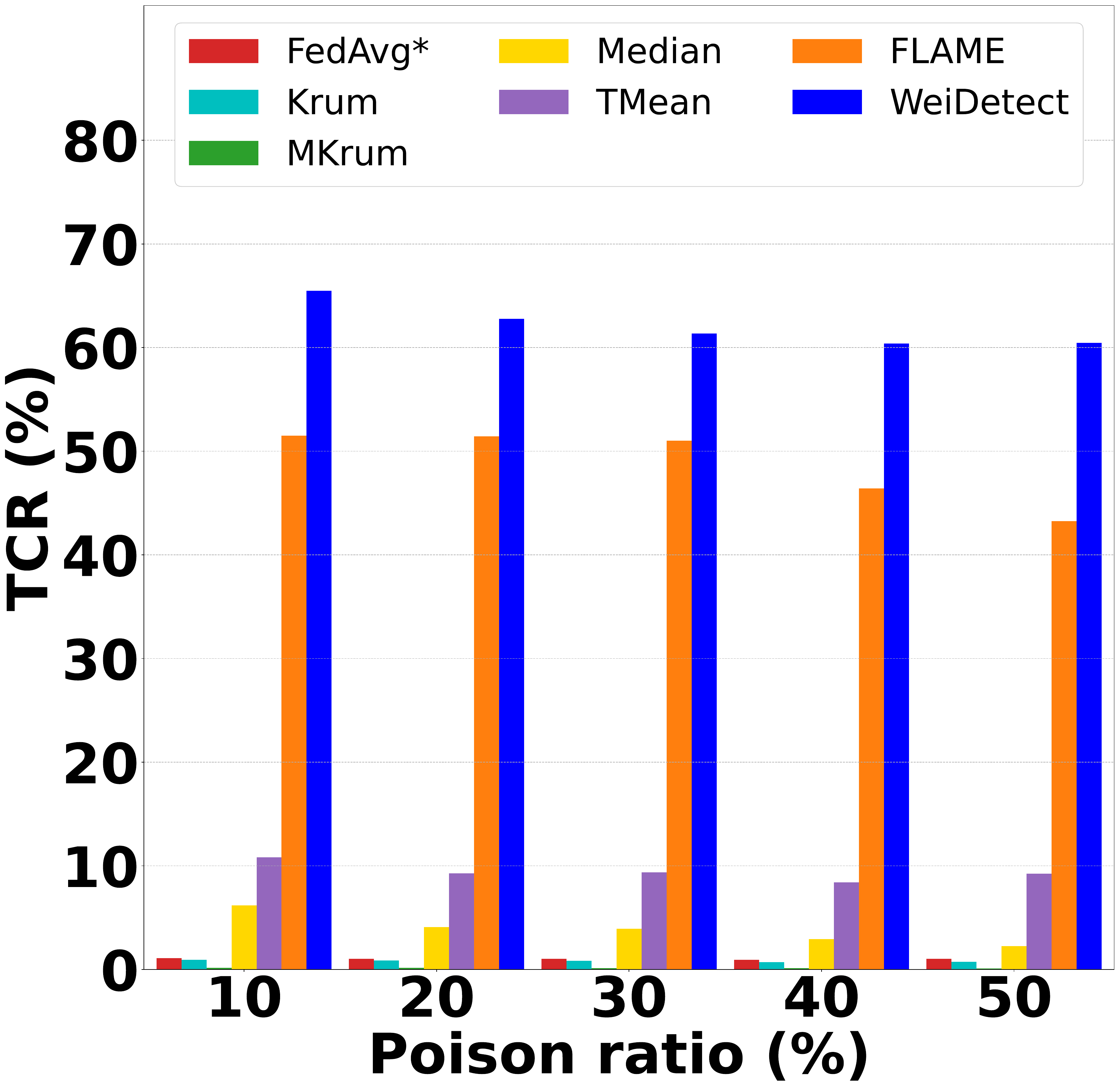} 
    \caption{GN}
    \label{fig:cicdarknet_sl_gn}
\end{subfigure}

\caption{Performance of WeiDetect on the CIC-Darknet2020 dataset under single targeted class adversarial attacks, compared with other approaches.}
\label{fig:darknet_sl_TCR}
\end{figure*}

\begin{table*}[h!]
\centering
\caption{Performance comparison of WeiDetect SOTA methods against triple target class adversarial attack in CIC-Darknet2020 dataset. The best score is highlighted in bold.}
\label{tab:tl_cicdarknet}
\footnotesize
\resizebox{\textwidth}{!}{
\begin{tabular}{ccccccccc}
\toprule

Attack strategy& Poison ratio$\downarrow$& FedAvg* & Krum & MKrum & Median  & TMean& FLAME & WeiDetect \\
\toprule

\multirow{5}{*}{FGSM} &10\% & 86.132& 86.307&86.130  & 82.164   &84.566& 82.177 &  \textbf{89.687} \\

&20\% & 86.0 &86.218   &87.642   &
81.223  & 84.302 &88.049   &  \textbf{89.653}  \\
&30\% & 86.005  &86.217   & 87.500  & 80.898 &  84.378& 88.042  &  \textbf{89.716}  \\
&40\% &   86.020 &  86.189 & 87.947  & 81.135 &84.280  &  87.843 &  \textbf{89.638}  \\
&50\% & 85.933  & 86.144  &  87.692 & 80.915 & 84.158 & 86.730  &  \textbf{89.674}  \\ 
\hline
\multirow{5}{*}{PGD} &10\%  &85.995& 86.307&88.128  & 81.705   &84.349&  82.347&  \textbf{88.471} \\
&20\% &  85.787 & 86.218  &  87.677 & 81.251 & 83.600 & 83.416  &  \textbf{ 88.009}  \\ 
&30\% & 85.378  & 86.217  &   87.402& 81.444 &83.298  & 84.216  &  \textbf{ 88.319}\\
&40\% &  85.447 & 86.189  & 86.712  & 80.655 & 82.521 & 83.365  &  \textbf{88.523 }  
\\ 
&50\%  & 85.362 &86.144   & 87.001  & 80.560 & 82.564 & 83.114  &  \textbf{87.889 }  \\ 

\hline
\multirow{5}{*}{GN} &10\% & 85.805&85.908 &87.687  & 81.472   &84.231& 83.492 &  \textbf{89.373} \\
&20\% &85.427  & 85.958  &  86.968 &  81.010&83.522  & 82.756  &  \textbf{89.512 }  \\
&30\% & 85.126  & 85.718  & 86.925  & 80.472 &83.160  &  82.973 &  \textbf{89.395 }  \\
&40\% & 85.047  & 85.668  &  86.614 &80.140  &82.521  &  82.866 &  \textbf{ 89.169}  \\
&50\% & 84.677 & 8.568  &86.374   & 81.339 & 81.550 & 83.468  &  \textbf{89.335 }  \\ 
\bottomrule
\end{tabular}}
\end{table*}

\begin{figure*}[h!]
\centering
\begin{subfigure}{0.33\textwidth}
    \includegraphics[width=\linewidth]{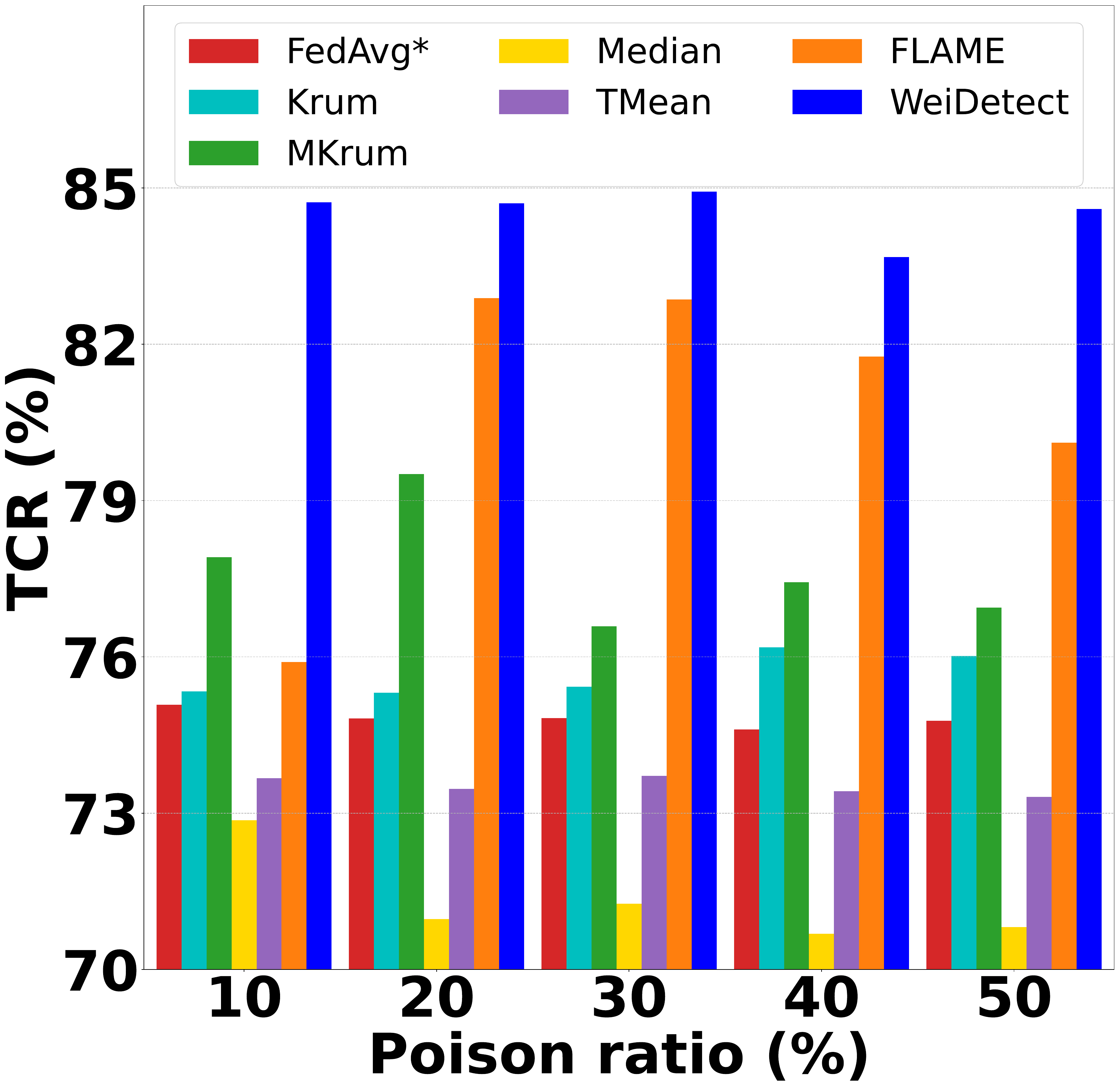} 
    \caption{FGSM}
    \label{fig:cicdarknet_tl_fgsm}
\end{subfigure}
\hfill
\begin{subfigure}{0.33\textwidth}
    \includegraphics[width=\linewidth]{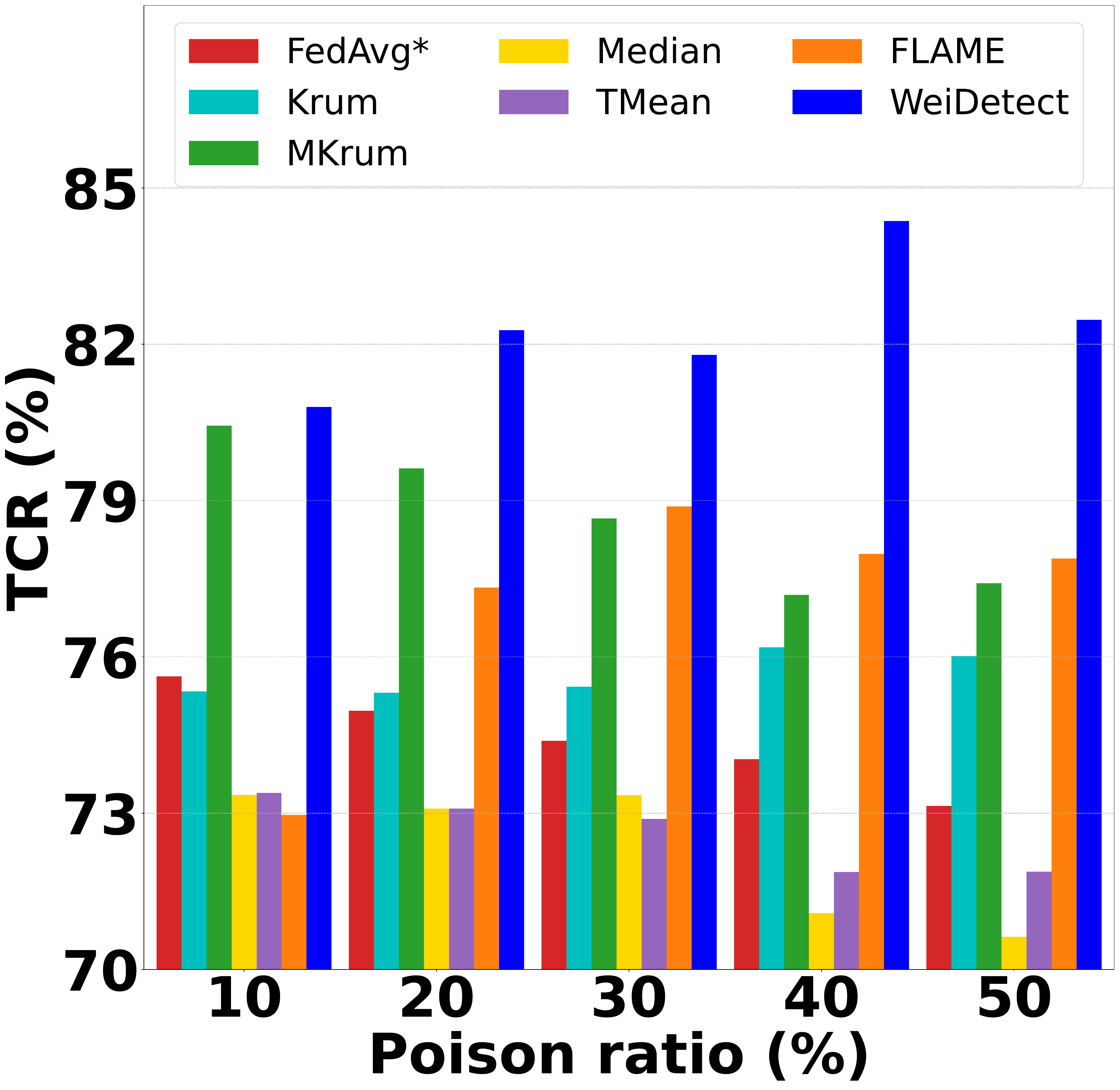} 
    \caption{PGD}
    \label{fig:cicdarknet_tl_pgd}
\end{subfigure}
\hfill
\begin{subfigure}{0.33\textwidth}
    \includegraphics[width=\linewidth]{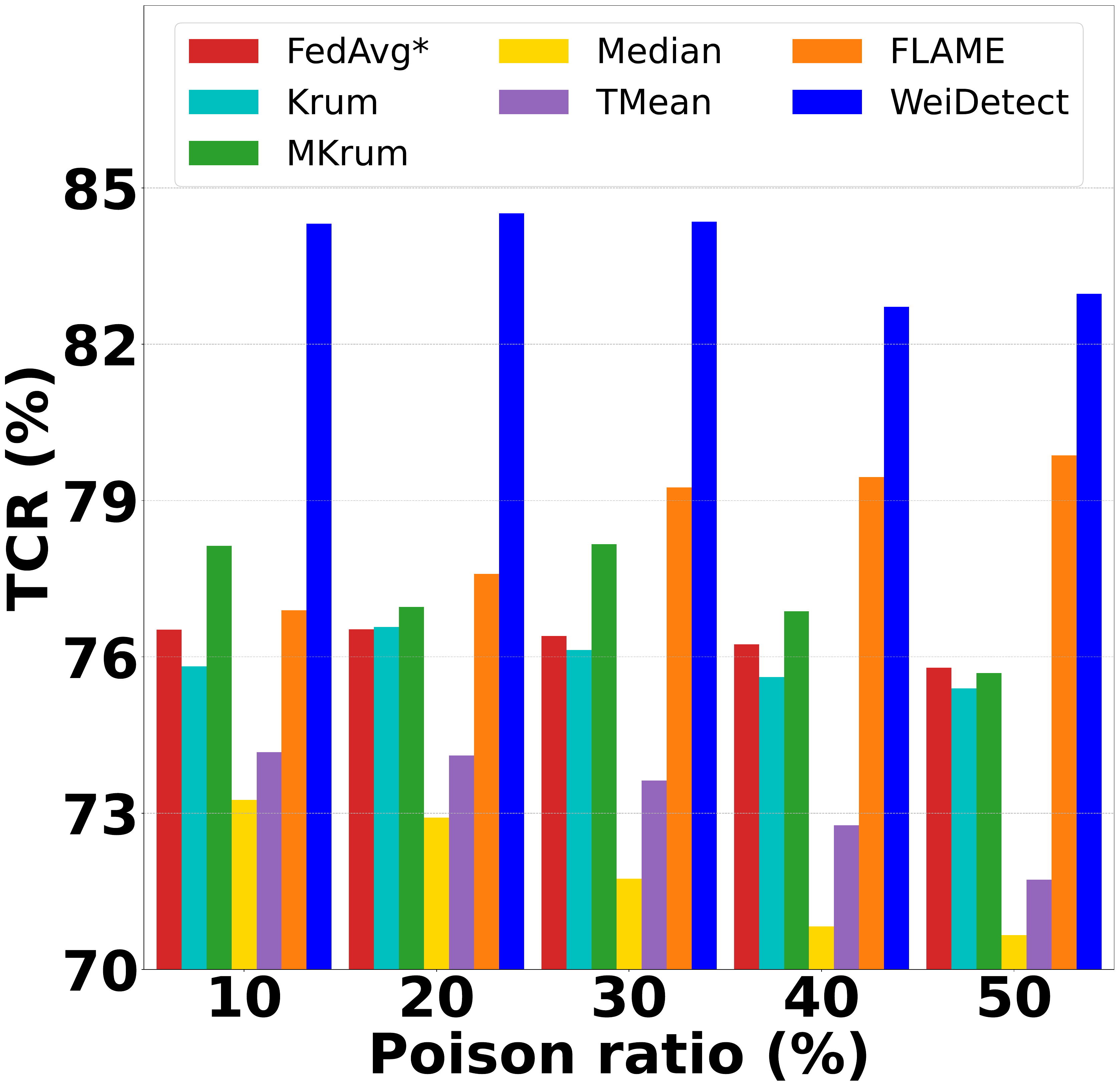} 
    \caption{GN}
    \label{fig:cicdarknet_tl_gn}
\end{subfigure}

\caption{Target class recall performance of WeiDetect on the CIC-Darknet2020 dataset, with the attacker focusing on a triple-label attack, compared to other defense approaches}
\label{fig:darknet_tl_TCR}
\end{figure*}

\begin{table*}[h!]
\centering
\caption{Performance comparison of WeiDetect with SOTA methods against single target class adversarial attack in CSE-CIC-IDS2018 dataset. }
\label{tab:sl_csecicids2018}
\footnotesize
\resizebox{\textwidth}{!}{
\begin{tabular}{ccccccccc}
\toprule

Attack strategy& Poison ratio$\downarrow$& FedAvg* & Krum & MKrum & Median  & TMean& FLAME & WeiDetect \\
\toprule

\multirow{5}{*}{FGSM} &10\% &91.784 &92.747 &92.664& 87.576  &88.816 & 93.170 &93.283 \\

&20\% & 91.774&92.705 &92.665  &  87.567  &88.828&93.212 &  93.213 \\

&30\% &91.779 &92.696 & 92.665 & 87.560   &88.879& 93.224&  93.225 \\

&40\% &91.774 &92.700 & 92.664 &  87.501  &88.896& 93.192&  93.193 \\

&50\% &91.763 & 92.576&92.660  & 87.495   &88.898&93.407 &  93.176 \\

\hline

\multirow{5}{*}{PGD} &10\%  & 91.827&92.756 & 93.863&87.637 & 88.803   &93.123&93.295\\
&20\% &91.461 &92.741 &93.864 & 87.530   &88.779& 93.140&  93.334 \\
&30\% &91.859& 92.737& 93.874&87.435  &88.768    &93.570 &  93.402 \\
&40\% &91.860&92.724 &93.858 &87.377  & 88.762   &94.807 &  93.444 \\
&50\%  &91.863 &92.677 & 93.873 & 87.248   &88.747& 93.356&  93.453 \\

\hline
\multirow{5}{*}{GN} &10\% &91.819 & 92.935& 92.695 & 87.479   &88.708&93.510 &  93.283 \\

&20\% &91.834 &92.990 & 92.746 &  86.979  &88.660& 93.298&  93.451 \\

&30\% & 91.833&92.713 &92.755  & 86.307&88.568&93.066&  93.265 \\

&40\% & 91.832&92.702 & 92.744 & 84.860 &88.446&92.966 &  92.904 \\

&50\% &91.831 &92.727&92.784 &84.186 & 88.308 &92.801& 92.817 \\
\bottomrule
\end{tabular}}
\end{table*}

\begin{figure*}[h!]
\centering
\begin{subfigure}{0.33\textwidth}
    \includegraphics[width=\linewidth]{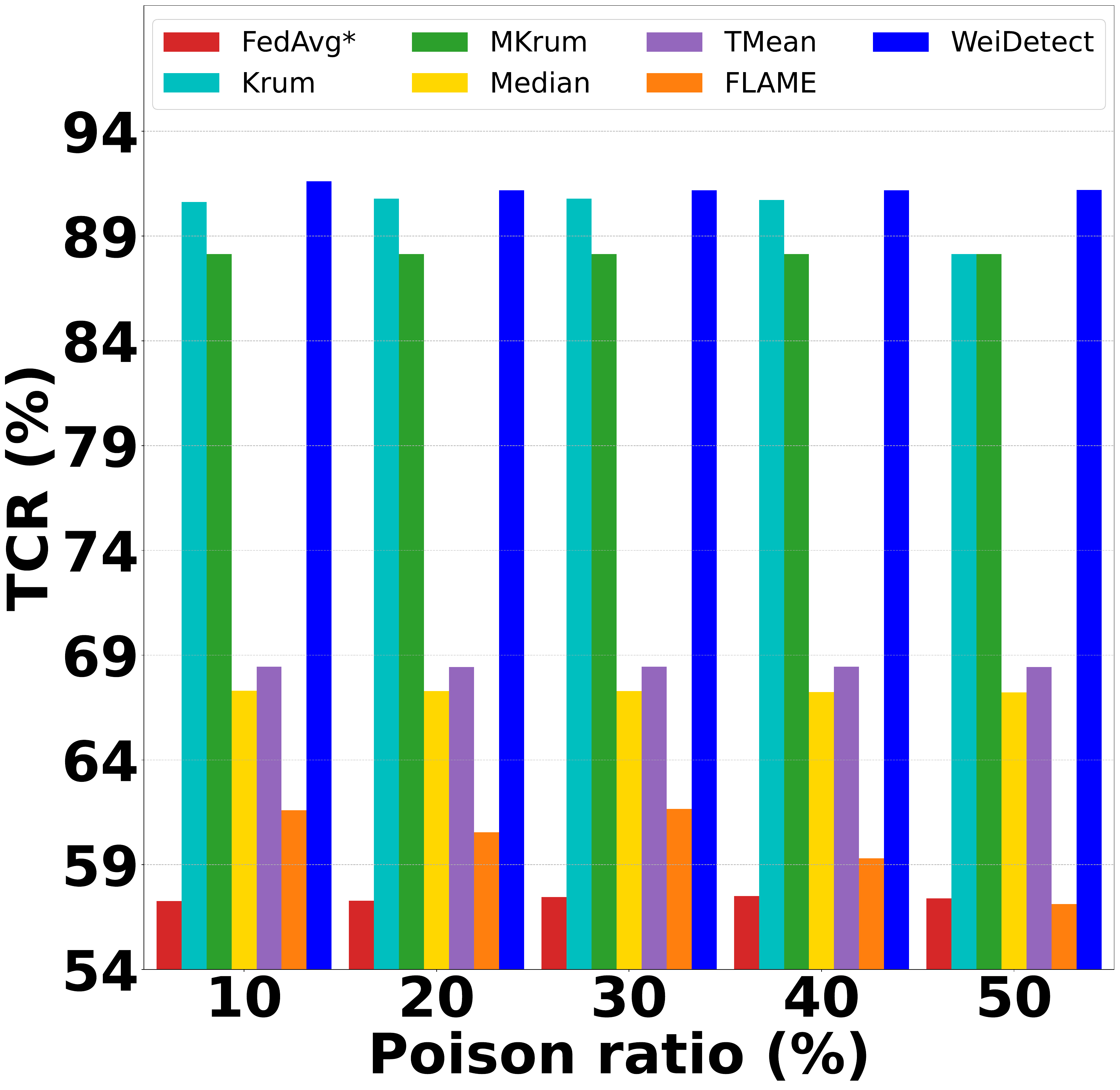} 
    \caption{FGSM}
    \label{fig:cicids_sl_fgsm}
\end{subfigure}
\hfill
\begin{subfigure}{0.33\textwidth}
    \includegraphics[width=\linewidth]{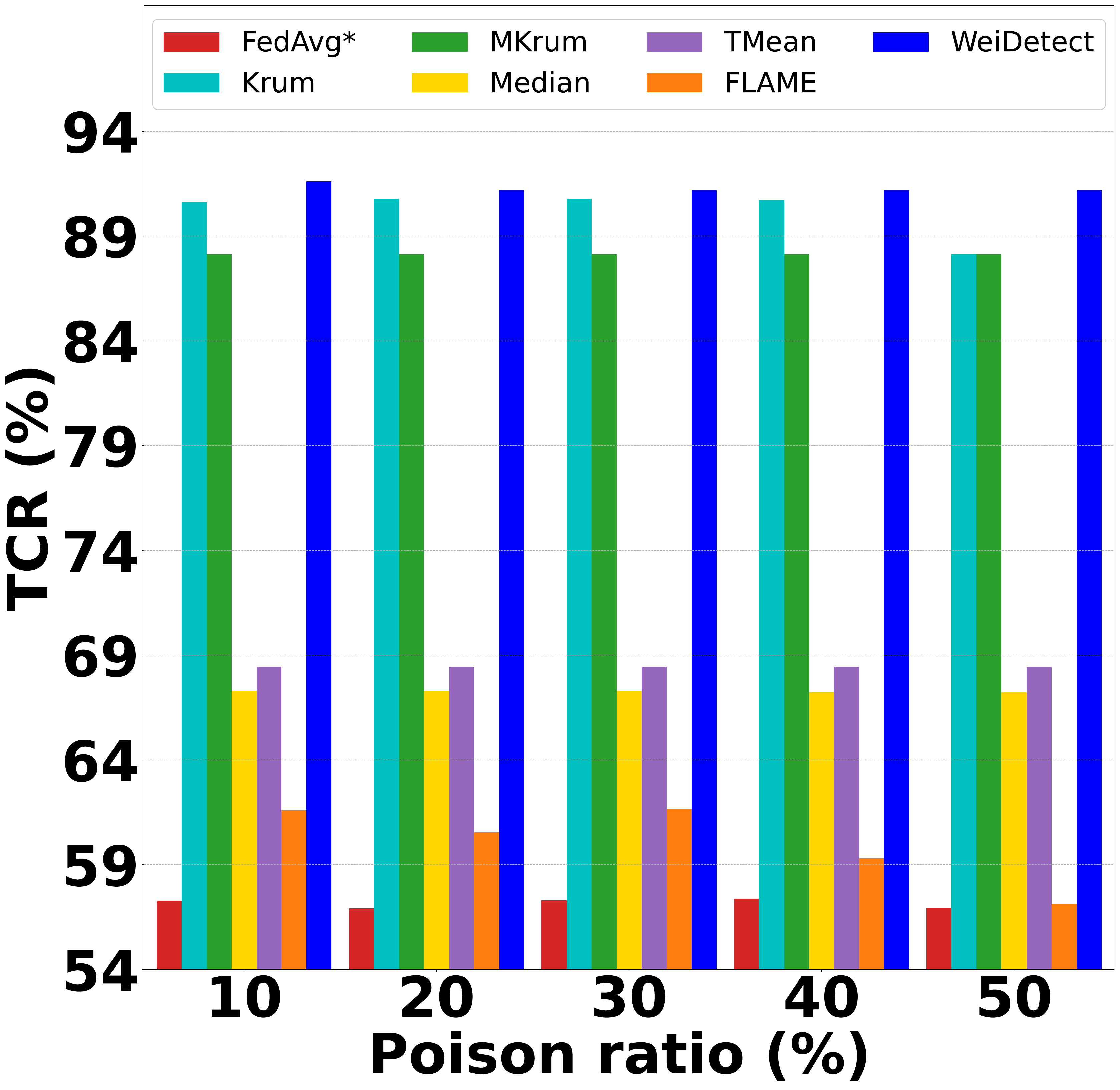} 
    \caption{PGD}
    \label{fig:cicids_sl_pgd}
\end{subfigure}
\hfill
\begin{subfigure}{0.33\textwidth}
    \includegraphics[width=\linewidth]{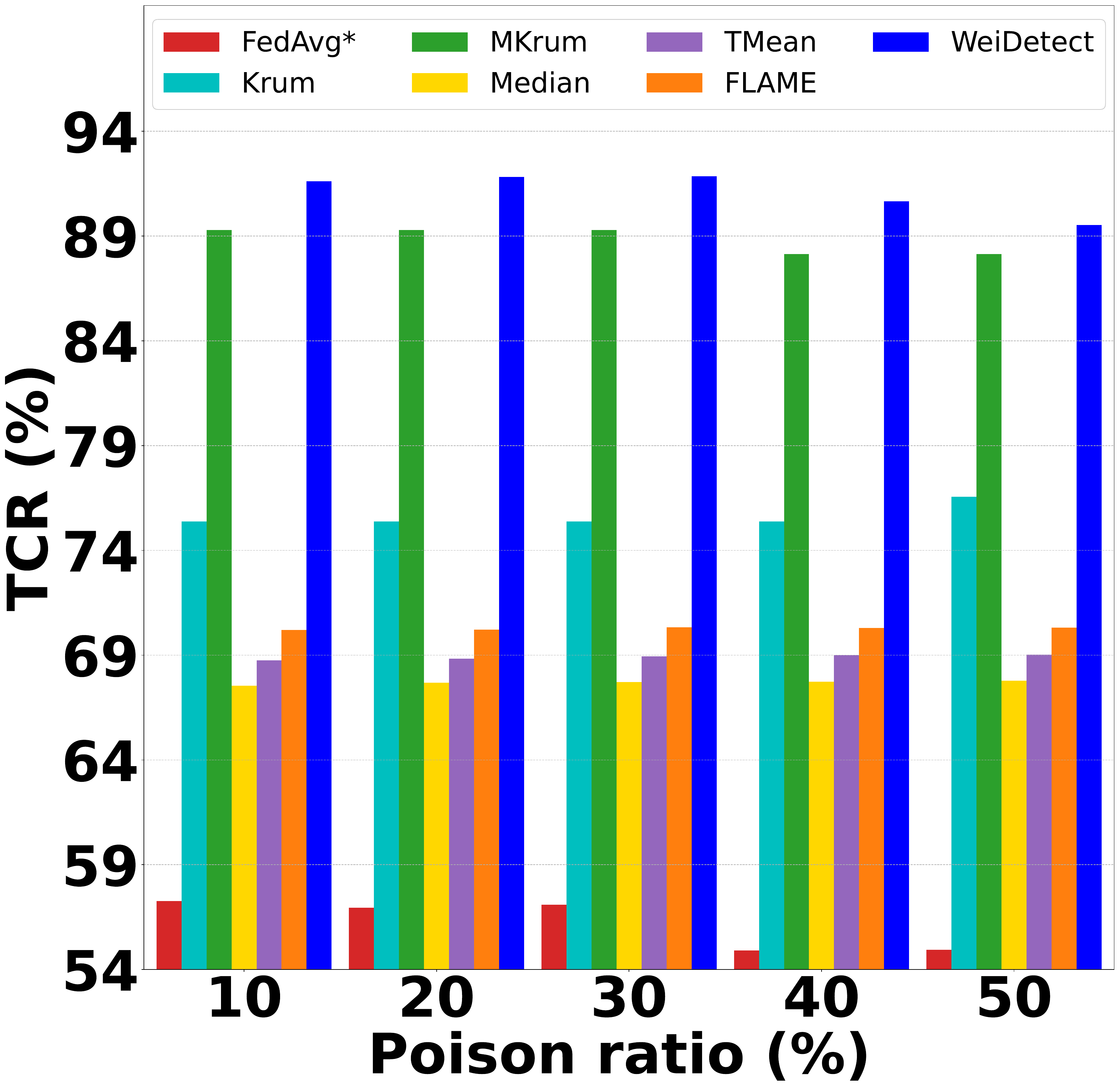} 
    \caption{GN}
    \label{fig:cicids_sl_gn}
\end{subfigure}

\caption{Target class recall performance of WeiDetect on the CSE-CIC-IDS2018 dataset, with the attacker focusing on a single-label attack, compared to other defense approaches}
\label{fig:cicids_sl_TCR}
\end{figure*}

\begin{table*}[ht]
\centering
\caption{Performance comparison of WeiDetect with SOTA methods against triple target class adversarial attack in CSE-CIC-IDS2018 dataset. The best score is highlighted in bold.}
\label{tab:tl_csecicids2018}
\footnotesize
\resizebox{\textwidth}{!}{
\begin{tabular}{ccccccccc}
\toprule

Attack strategy& Poison ratio$\downarrow$& FedAvg* & Krum & MKrum & Median  & TMean& FLAME & WeiDetect \\
\toprule

\multirow{5}{*}{FGSM} &10\% &91.705 &92.363 & 92.344   &81.609&84.632 & 93.933& \textbf{93.493} \\

&20\% &91.682 &92.336 & 92.415& 81.648&84.811&92.450 &\textbf{93.438} \\

&30\% &91.683 &92.318 &  92.620& 81.159 &84.776& 92.810&  \textbf{93.495} \\

&40\% & 91.676&92.282 &92.600  & 81.753   &85.006&92.508&\textbf{93.471} \\
&50\% &91.675 &92.205 & 92.390 & 80.927 &84.993& 92.237&  \textbf{93.445} \\
\hline

\multirow{5}{*}{PGD} &10\%  & 91.912&92.499  & 92.697   &91.974&92.245 &  92.577&\textbf{93.395} \\

&20\% &92.313 &92.460 & 92.695 &  91.989  &92.268& 92.121&  \textbf{93.331} \\

&30\% &92.031&92.446 & 92.634&  91.973&  92.283  & 92.111&  \textbf{93.163} \\

&40\% &91.463& 92.435&92.663 &91.984  &  92.283  &93.239 &  \textbf{93.495} \\

&50\%  &92.139 &92.420 & 92.600 & 91.979&92.288& 92.625&\textbf{93.219} \\

\bottomrule
\end{tabular}}
\end{table*}
\begin{figure*}[h!]
\centering
\begin{subfigure}{0.4\textwidth}
    \includegraphics[width=\linewidth]{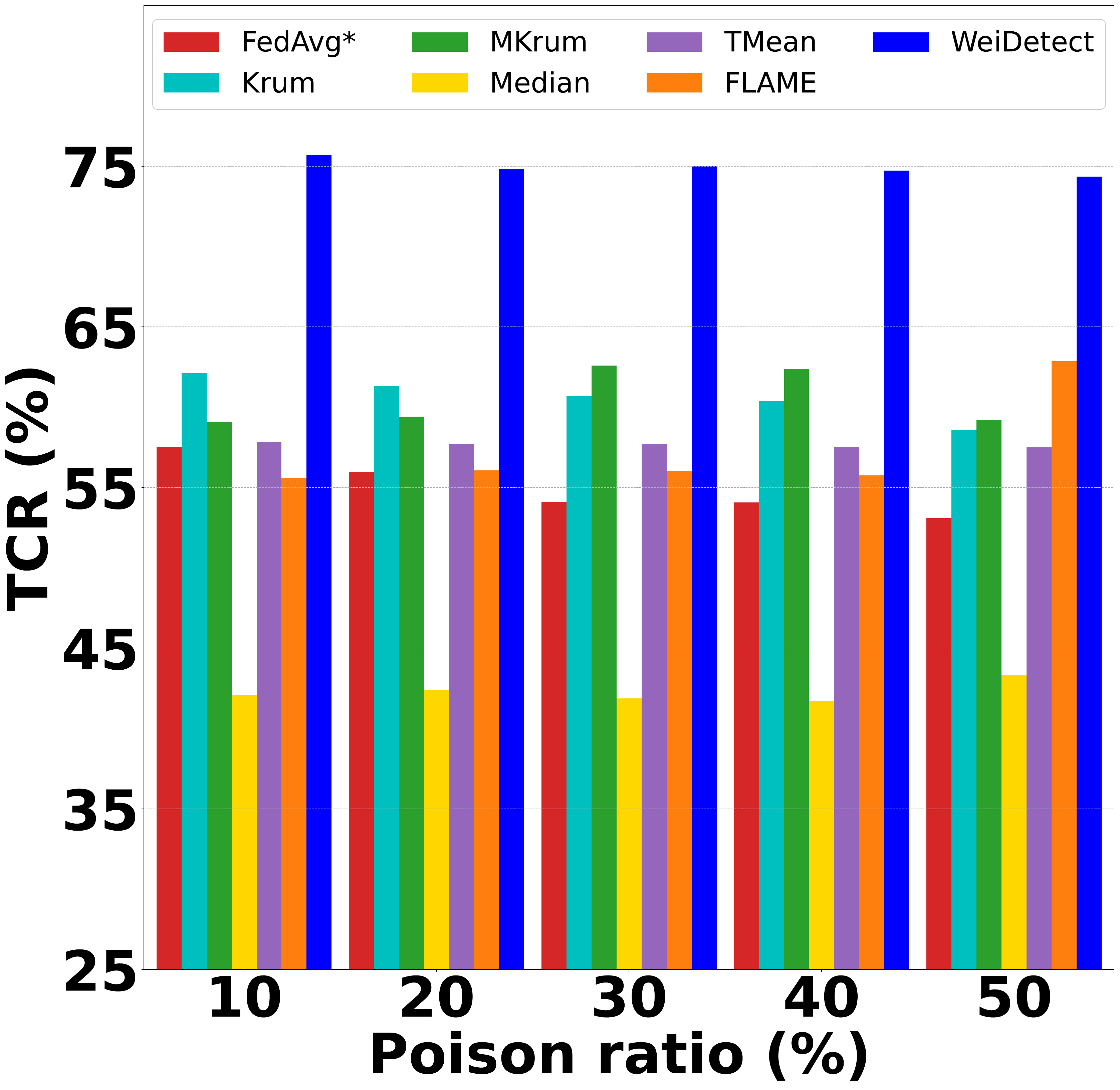} 
    \caption{FGSM}
    \label{fig:cicids_tl_fgsm}
\end{subfigure}
\begin{subfigure}{0.4\textwidth}
    \includegraphics[width=\linewidth]{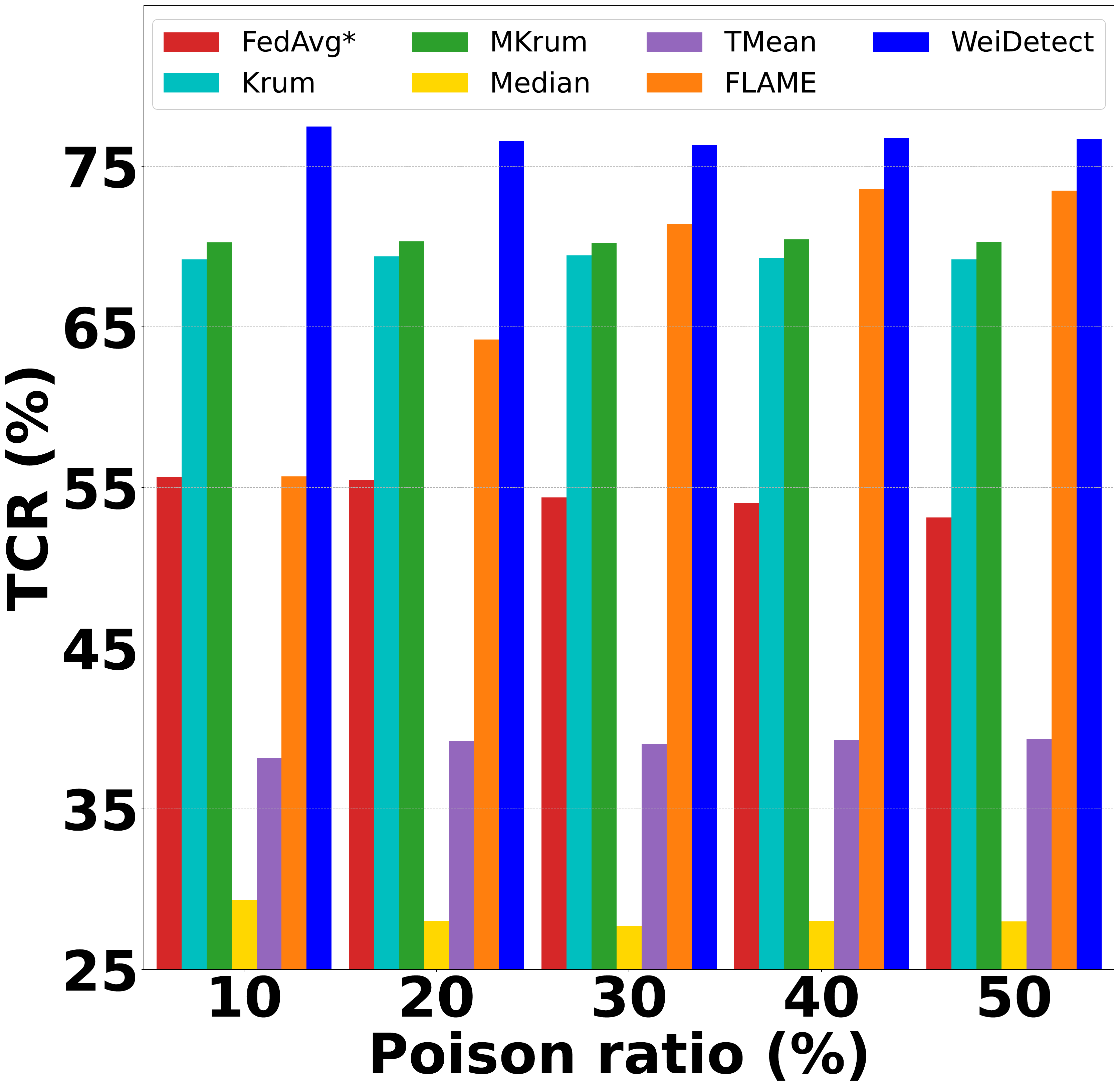} 
    \caption{PGD}
    \label{fig:cicids_tl_pgd}
\end{subfigure}

\caption{Target class recall performance of WeiDetect on the CSE-CIC-IDS2018 dataset, with the attacker focusing on a triple-label attack, compared to other defense approaches}
\label{fig:cicids_tl_TCR}
\end{figure*}
\subsubsection{Effectiveness of WeiDetect against data poisoning attack}

To demonstrate the effectiveness of our proposed WeiDetect against poisoning attacks on imbalanced data, we conducted experiments involving 15\% malicious participants executing the attack. Figure \ref{fig:darknet_sl_TCR} illustrates the performance comparison between WeiDetect and other approaches, regarding targeted class recall under different percentages of perturbed samples, in adversary-controlled participants for the CIC-Darknet2020 dataset. Additionally, Table \ref{tab:sl_cicdarknet} presents the global model F1-score for different poisoning ratios on CIC-Darknet2020 dataset.
The results demonstrate that, with 10\% of poisoned samples generated using the FGSM approach, the targeted class recall dropped significantly from 76.102\% to 3.588\% in FedAvg (refer Figure \ref{fig:cicdarknet_sl_fgsm}), which had a global model F1-score of 75.965\%, compared to the baseline approach without any defense. In this scenario, the empirical findings indicate that our proposed approach outperforms other methods. At the same time, the target class recall of state-of-the-art methods drops significantly. WeiDetect, on the other hand, effectively identifies malicious clients and mitigates the attack, achieving a target class recall of 74.631\% with an F1-score of 87.014\%. This shows more than 73\% improvement in target class recall and a 13\% improvement in the global model F1-score compared to the other state-of-the-art approaches.

As the poisoning percentage increases, the impact on model performance becomes more pronounced. In the case of Krum and MKrum, the TCR of the targeted class drops to 0 when the poisoning percentage reaches 30\%. In contrast, WeiDetect maintains high performance even with 50\% of poisoned samples, achieving a TCR of 67.628\% and a global model F1-score of 88.108\%. This results in a more than 67\% improvement in TCR and a 14\% improvement in global model performance compared to other approaches.

We also evaluate the effectiveness of the proposed defense model against other adversarial sample generation approaches, such as PGD (Figure \ref{fig:cicdarknet_sl_pgd}) and GN (Figure \ref{fig:cicdarknet_sl_gn}). The experimental results show that WeiDetect outperforms the other existing approaches in every adversarial attack scenario.
For extreme attack scenarios in our setting, the PGD attack results indicate that our approach achieves approximately 11\% improvement in the F1-score and 65\% improvement in targeted class accuracy. In the PGD attack scenario, Krum and MKrum exhibit a source class recall close to 0, even with 10\% poisoned samples. A similar trend is also observed in the GN adversarial attack scenario. This behavior is due to both approaches performing poorly on non-IID data distributions, likely because they rely on strong assumptions about the number of malicious participants and aggressively remove outliers using Euclidean distance, which likely leads to discarding benign updates.
Additionally, our method performs well across all metrics for the GN adversarial case, showing more than 17\% to 60\% improvement in target class recall and approximately 14\% improvement in the F1-score. A similar trend is also observed in the PGD attack scenarios. Furthermore, under the targeted attack, all the existing approaches maintain the F1-score of the global model while significantly reducing the TCR with a higher percentage of poisoned samples.

To further assess the performance of WeiDetect, we investigate an attack scenario where the attacker targets multiple classes to induce poisoned samples, with a specific focus on triple-class poisoning. As shown in Figure \ref{fig:darknet_tl_TCR} and Table \ref{tab:tl_cicdarknet}, the results demonstrate that the proposed approach outperforms all other methods across all evaluation metrics for the CIC-Darknet2020 dataset. Specifically, the F1-score of the global model improved by approximately 2–9\%, 0.888-9\%, and 2–8\% under the FGSM, PGD, and GN attack strategies, respectively. Similarly, for the TCR, notable improvements of 14\%, 12\%, and 13\% were observed in the extreme attack scenario (where 50\% of the samples are poisoned) for FGSM, PGD, and GN, respectively.

In the case of the CSE-CIC-IDS2018 dataset, we observed similar trends. Table \ref{tab:sl_csecicids2018} presents the global model F1-score achieved by the different approaches when the attacker focuses on a single targeted class. The results demonstrate that the proposed approach outperformed other methods in most cases. Additionally, the global model performance of the FLAME approach is comparable to that of WeiDetect. However, the target class recall for WeiDetect shows superior results compared to other methods. Figure \ref{fig:cicids_sl_TCR} visually represents these findings. We observed that WeiDetect outperformed state-of-the-art approaches, maintaining a TCR above 90\% and achieving an F1-score of 93\% approximately for the global model. Table \ref{tab:tl_csecicids2018} and Figure \ref{fig:cicids_tl_TCR} present the performance evaluation of WeiDetect against state-of-the-art techniques in mitigating triple-target class adversarial attacks. A comparable pattern emerges, showcasing WeiDetect’s effectiveness over existing methods by maintaining approximately TCR above 75\% and achieving above 93\% F1-score for the global model.

\par
\begin{figure*}[h!]
\centering
\begin{subfigure}{0.4\textwidth}
    \includegraphics[width=\linewidth]{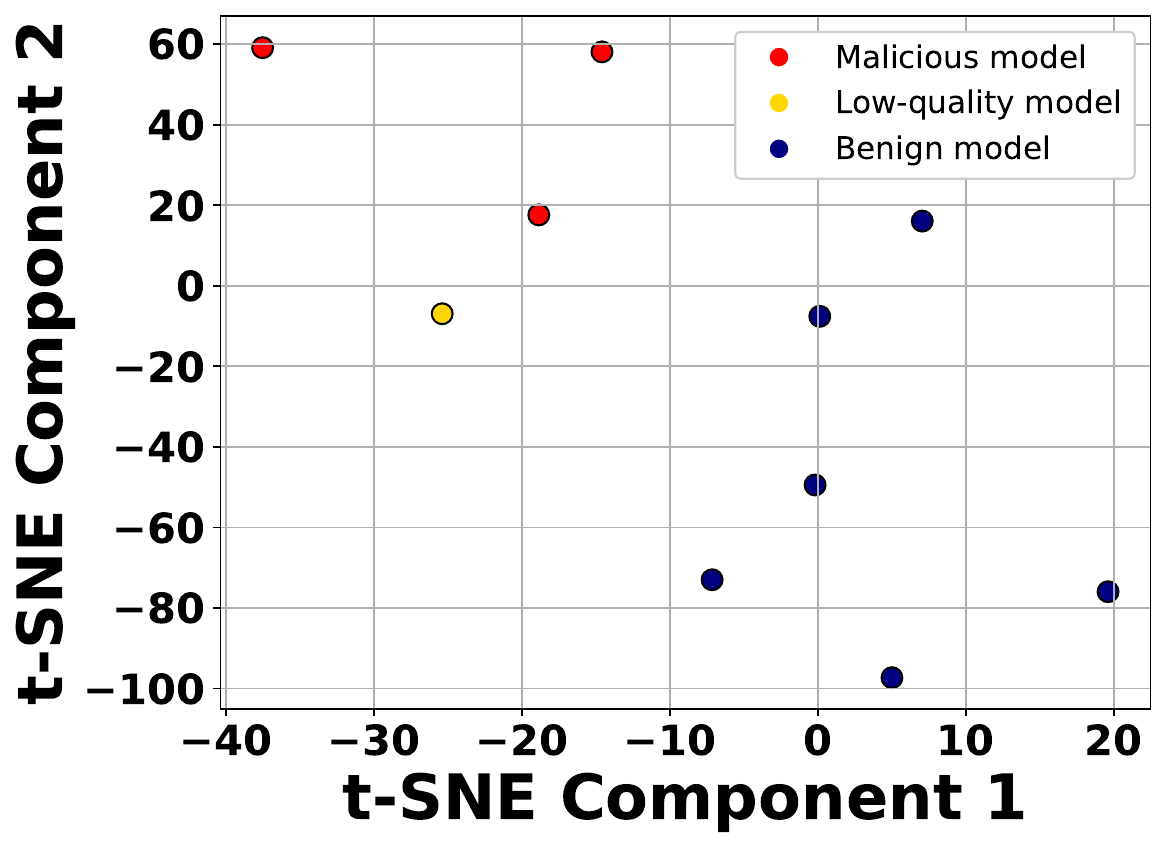} 
    \caption{Middle round }
    \label{fig:TSNE_MID}
\end{subfigure}
\hfill
\begin{subfigure}{0.4\textwidth}
    \includegraphics[width=\linewidth]{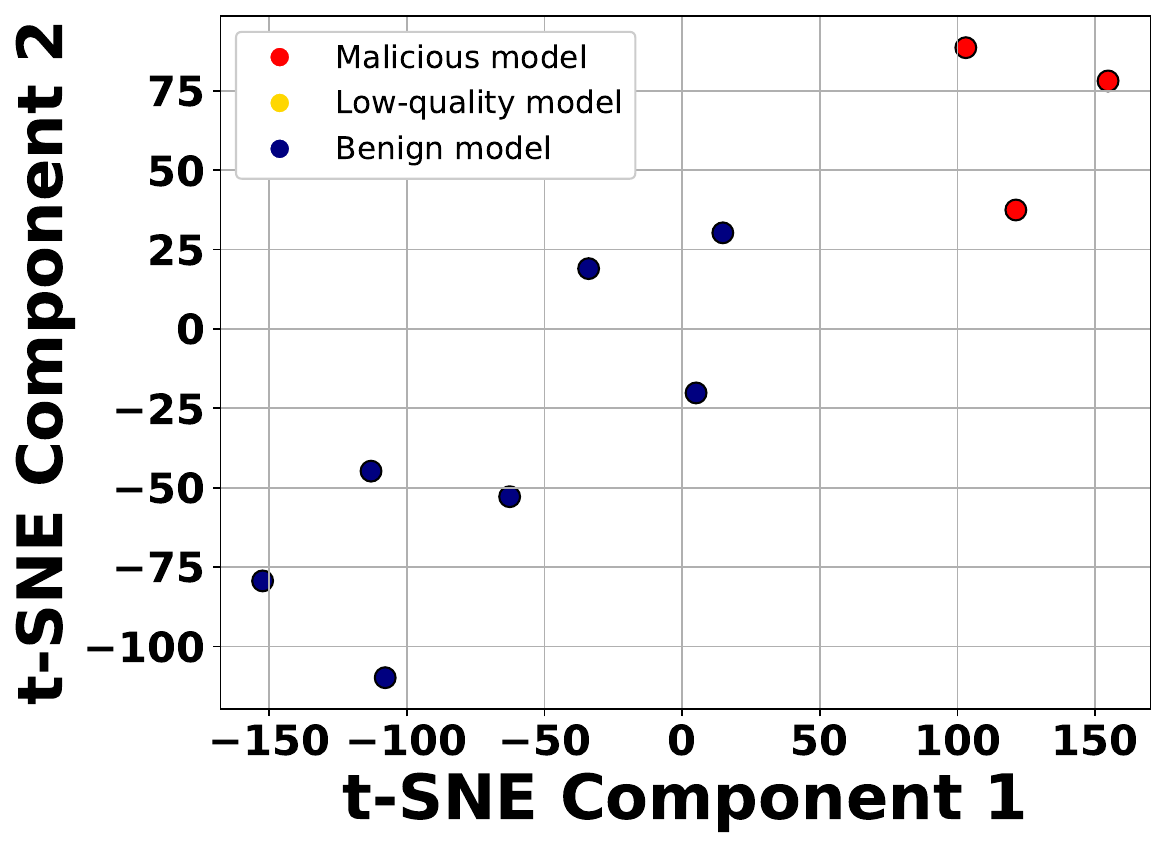} 
    \caption{End Round}
    \label{fig:TSNE_END}
\end{subfigure}

\caption{The effectiveness of the proposed approach, WeiDetect, in distinguishing adversarial updates from benign updates across different communication rounds.}
\label{fig:TSNE}
\end{figure*}

 To enhance the visual representation of our findings, we employed t-SNE visualization in the middle and final training rounds of the WeiDetect. Figure \ref{fig:TSNE} illustrates the effectiveness of our approach in classifying adversarial, low-quality, and benign updates. In the middle rounds, WeiDetect detects malicious and low-quality models, which exhibit similar properties and are represented by red and yellow in the figure. By the final rounds, WeiDetect successfully identified malicious models and distinguished them from benign ones. As rounds increase, benign models (including low-quality models in the preliminary rounds) contribute to overall performance improvement, enabling a clearer separation of malicious models. For the t-SNE plot, we used results obtained from both phases to demonstrate the efficiency of our approach.

\subsection{Effectiveness of WeiDetect against label-flipping attack}
To evaluate WeFighter's effectiveness in the FL-based NIDS, we conducted additional experiments to assess its robustness against a Label Flipping (LF) attack. In this attack, the adversary flips the labels of the victim class (target class $C_s$ ) to other classes (poisoned class $C_p$), represented as $C_s \rightarrow C_p$. Our work targets the most misclassified label combinations from the baseline scenario to adjust the class assignments. Unlike more sophisticated poisoning methods, LF is particularly advantageous for adversaries with limited knowledge of the FL-based NIDS, as it is simple to implement and incurs minimal computational overhead. Despite its simplicity, LF poses a significant threat to FL-based NIDS, causing severe degradation in model performance and compromising the system's ability to detect intrusions accurately.

\begin{table*}[ht]
\centering
\caption{Performance comparison of WeiDetect with SOTA methods under label flipping attack scenario using the F1-score of the global model.}
\label{tab:labelflipping}
\footnotesize
\resizebox{\textwidth}{!}{
\begin{tabular}{c c c c c c c c c c c c c}
\toprule

\multirow{2}{*}{Poison ratio$\downarrow$} & \multicolumn{6}{c}{Single label flipping attack} & \multicolumn{6}{c}{Triple label flipping attack} \\ 
\cmidrule(lr){2-7} \cmidrule(lr){8-13} 

& Krum & MKrum & Median  & TMean & FLAME & WeiDetect & Krum & MKrum & Median  & TMean & FLAME & WeiDetect \\
\midrule

\multicolumn{13}{c}{CSE-CIC-IDS2018} \\
\midrule

10\% & 93.502 & 93.776 &92.748&89.416  &95.210  &\textbf{95.934}  & 93.813 &93.777  & 83.204 &86.891  &94.984  & \textbf{95.176}   \\
20\% & 93.386 & 93.906 &92.697&89.794  &94.997  & \textbf{95.325} & 93.204 &94.062  & 81.617 &84.778  & 95.448 & \textbf{95.764 }  \\
30\% & 93.380 &92.033  &92.693& 89.888 &94.980  &\textbf{95.256}  &92.985  &93.895  &81.314  & 83.970 & \textbf{95.359} & 94.911   \\
40\% & 93.331 & 92.057 &92.682& 88.393 &94.651  &\textbf{95.333}  & 92.314 & 94.241 & 80.501 & 81.136 &94.424  &\textbf{94.968}    \\
50\% & 93.324 &92.120  &92.505& 86.780 &94.202  & \textbf{94.256} &92.227  &94.255  & 80.168 &80.827  & 94.431 &\textbf{ 94.630} \\
\midrule
\multicolumn{13}{c}{CICDarknet2020} \\
\midrule

10\% & 88.944 & 89.157 & 83.297 & 84.702 & 88.777 &\textbf{89.218 } & 89.171 & 89.033 & 83.654 & 84.880 & 89.156 &  \textbf{89.223}\\
20\% & 88.580 &89.145  &82.850  &84.075  & 88.346 &\textbf{89.197}    &  89.042&89.045  & 83.513 & 84.675 &83.274  & \textbf{89.079} \\
30\% & 88.104 & 88.318 &78.233  & 82.089 &82.285  &\textbf{89.200}   &88.733  &89.075  & 82.540 &84.448  & 84.022&\textbf{88.900} \\
40\% & 83.351 & 81.828 & 72.287 &74.182  &79.561  &\textbf{89.187} & 88.394 & 89.058 & 74.952 &83.258  & 83.021 & \textbf{88.798} \\
50\% &  76.321& 76.686 &72.321  &73.021  & 72.501 & \textbf{88.718} & 88.129 & 88.002 & 72.354 & 78.868 & 82.063 & \textbf{88.217} \\

\bottomrule
\end{tabular}}
\end{table*}

\begin{figure*}[h!]
\centering

\begin{subfigure}{0.245\textwidth}
    \includegraphics[width=\linewidth]{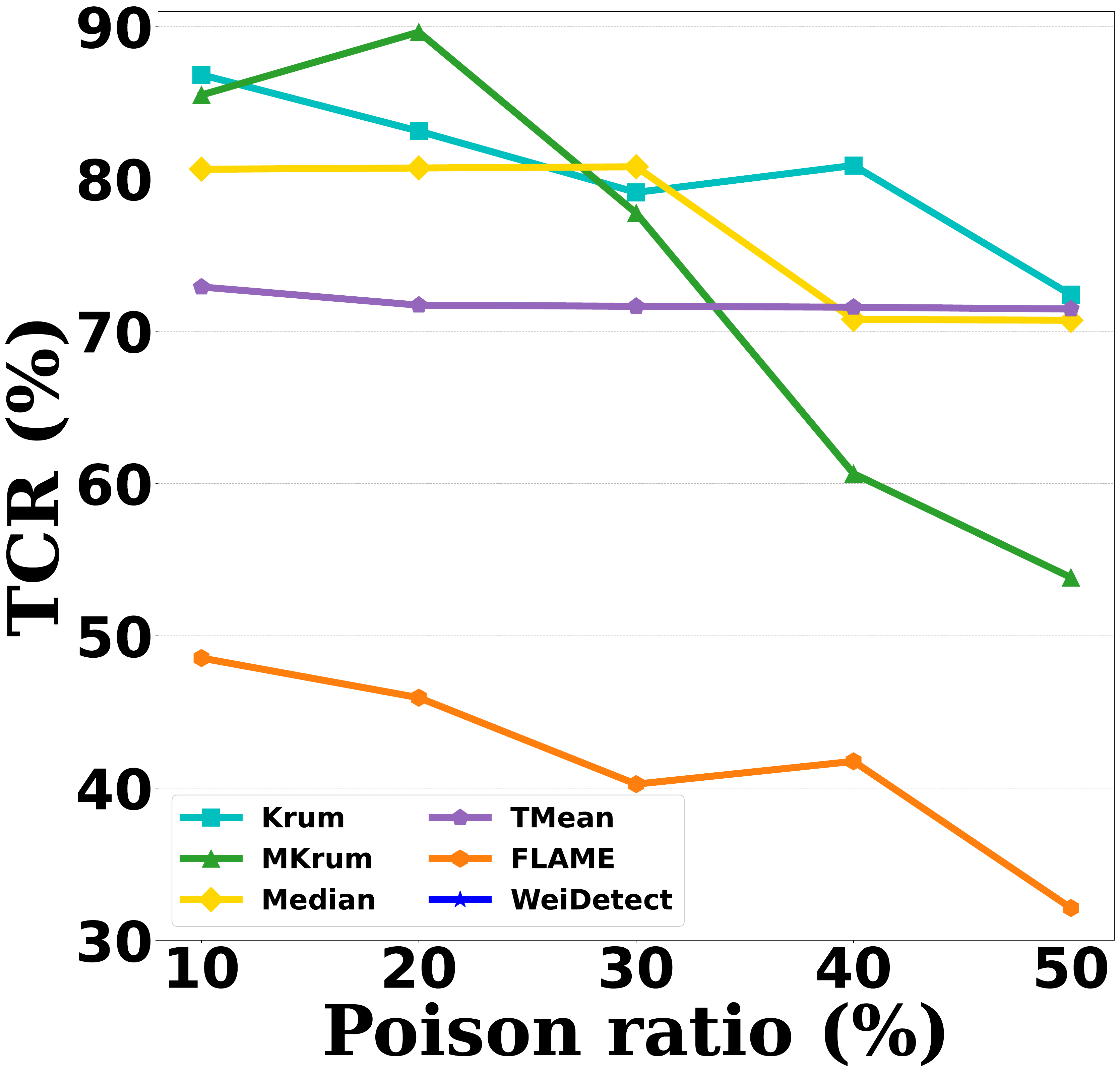} 
    \caption{CSE-CIC-IDS2018 (Single Label)}
    \label{fig:cicids_LF_SL}
\end{subfigure}
\hfill
\begin{subfigure}{0.245\textwidth}
    \includegraphics[width=\linewidth]{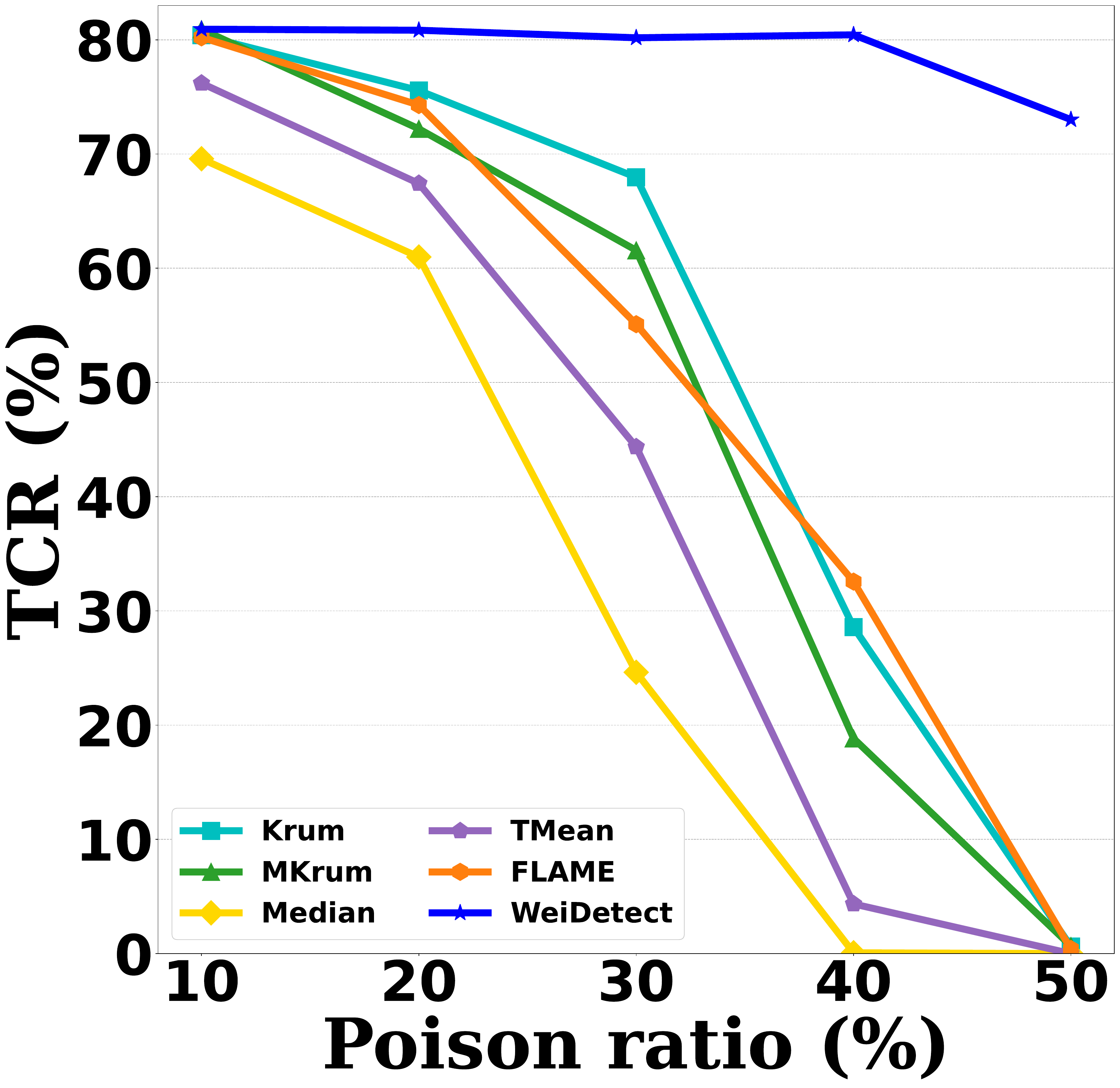} 
    \caption{CIC-Darknet2020 (Single Label)}
    \label{fig:darknet_LF_SL}
\end{subfigure}
\hfill
\begin{subfigure}{0.245\textwidth}
    \includegraphics[width=\linewidth]{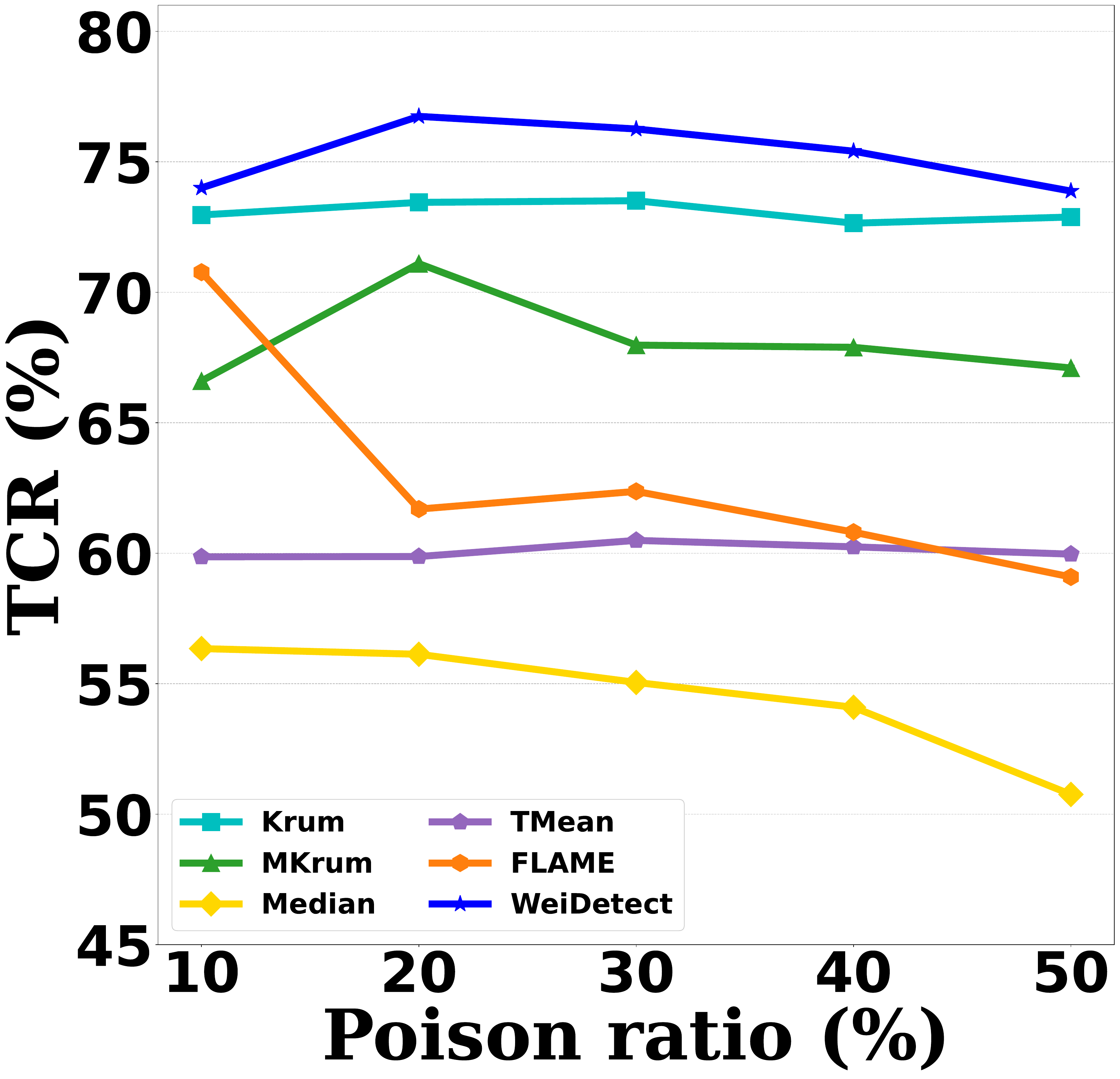}  
    \caption{CSE-CIC-IDS2018 (Triple Label)}
    \label{fig:cicids_LF_TL}
\end{subfigure}
\hfill
\begin{subfigure}{0.245\textwidth}
    \includegraphics[width=\linewidth]{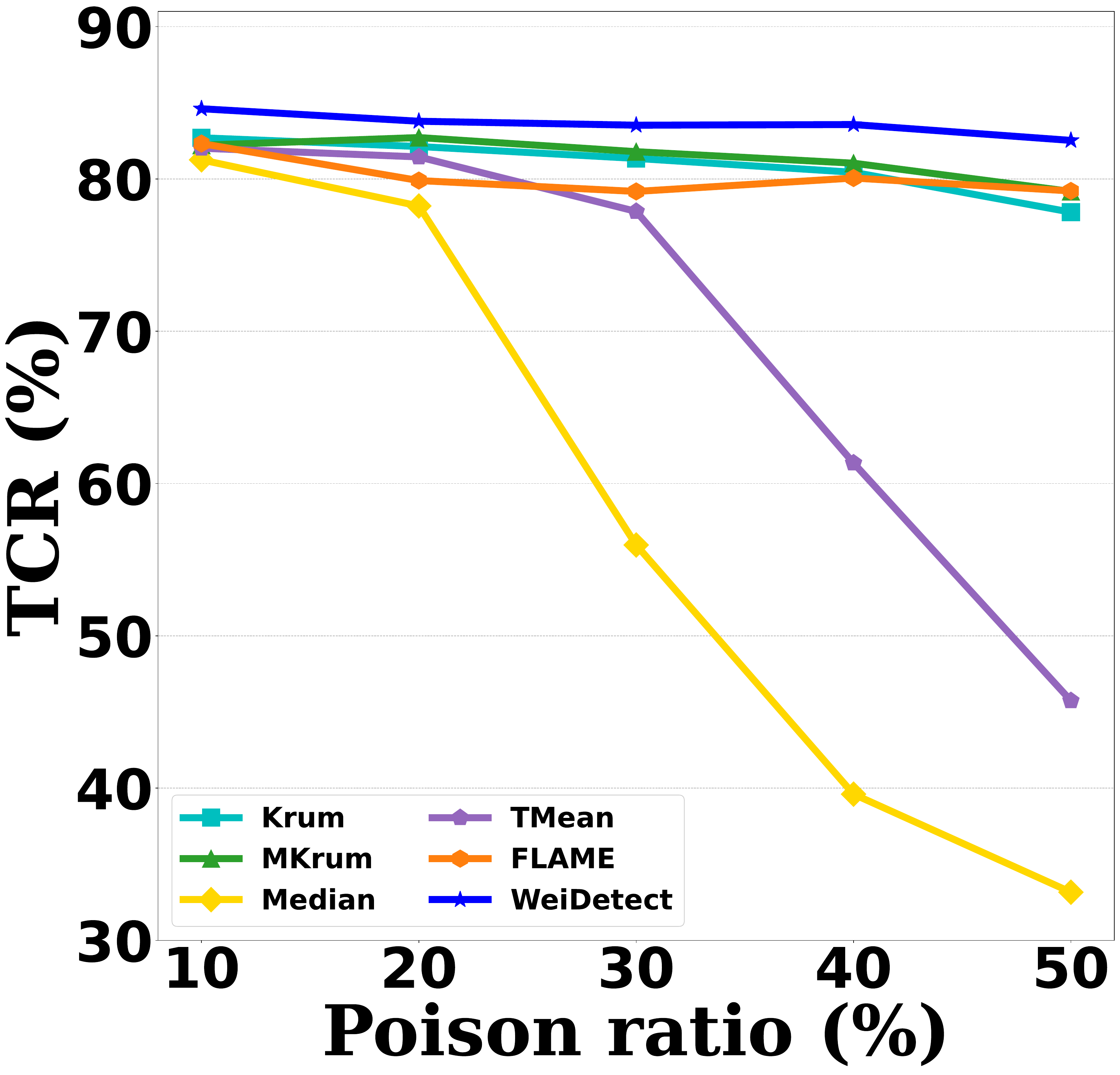}  
    \caption{CIC-Darknet2020 (Triple Label)}
    \label{fig:darknet_LF_TL}
\end{subfigure}
\hfill
\caption{Performance evaluation of WeiDetect using the class recall of the victim class, compared to other defense approaches against label-flipping attacks.}
\label{fig:LF_Attack}
\end{figure*}

For the LF attack setup, we simulate a scenario involving  $15\%$ malicious participants who flip the labels of the samples. We conducted experiments on both datasets with poisoned sample levels 10\%, 20\%, 30\%, 40\%, and 50\% within each compromised participant. Figure \ref{fig:LF_Attack} illustrates the performance of WeiDetect and other state-of-the-art approaches in this scenario. It also depicts the class recall of the victim class, represented as TCR. Additionally, Table \ref{tab:labelflipping} demonstrates the F1-score of the global model's performance under different levels of poisoned samples.  

As shown in the table, the proposed model outperforms all poisoning attack scenarios, achieving a high F1 score for the global model compared to other existing methods. For the CSE-CIC-IDS2018 dataset in the single-label attack scenario, WeiDetect maintained an F1-score of 95.934\% when the adversary poisoned 10\% of the samples. Additionally, it preserved a TCR value of 91.863\% (Figure \ref{fig:cicids_LF_SL}). Across all poisoning ratios, WeiDetect sustained a TCR above 91\%, outperforming Krum and MKrum by approximately 5\% to 6\% and surpassing FLAME, Median, and TMean by 11\% to 43\%. Krum maintained a stable TCR of around 86\% to 83\% at lower poisoning levels (10\% to 20\%) but declined to approximately 14\% as the poisoning ratio increased to 50\%. Similarly, MKrum exhibited a 32\% decline in TCR when 50\% of the samples were poisoned. However, both Krum and MKrum rely on prior knowledge of the number of malicious participants, limiting their adaptability in real-world NIDS deployments. In the CIC-Darknet2020 dataset, WeiDetect consistently outperforms other approaches across all poisoning levels. 

For the CIC-Darknet2020 dataset, in the single-label flipping attack scenario, we observed that WeiDetect, Krum, M-Krum, and FLAME exhibited comparable performance at lower poisoning intensities. However, as the poisoning intensity increased, the proposed approach maintained the F1-score of the global model at approximately 89\%. Additionally, WeiDetect preserved the TCR of the source class at 80\%.
Furthermore, the proposed approach outperformed other methods by approximately 12\% to 16\% when 50\% of the samples were poisoned by malicious clients. A similar trend was observed for the TCR values of the victim class. At this poisoning level, all other approaches resulted in zero recall for the victim class, whereas WeiDetect achieved a TCR closer to 73.012\% (Figure \ref{fig:darknet_LF_SL}). WeiDetect achieves this by leveraging validation-based scoring and Weibull distribution anomaly detection, identifying and excluding malicious participants or low-quality models during aggregation. These results highlight the effectiveness of WeiDetect in mitigating the impact of label-flipping attacks.

To further assess the resilience of our FL-based NIDS against poisoning attacks, we evaluated it under the triple-class attack scenarios. For CSE-CIC-IDS2018 dataset, the results depicted that WeiDetect effectively withstood high poisoning levels, maintaining an F1-score of 94.630\% and achieving a TCR value close to 75\%, outperforming other approaches (Figure \ref{fig:cicids_LF_TL}). For the CIC-Darknet2020 dataset, WeiDetect improved the F1-score by more than 16\% compared to other approaches and increased TCR values by approximately 49\% over competing methods (Figure \ref{fig:darknet_LF_TL}). Our study depicted minimal impact on the global model's F1-score, but in targeted attacks, the class recall of the victim class dropped significantly in other approaches.


\section{Discussion}
\label{sec:discussion}

\subsection{Computational overhead of WeiDetect}
To evaluate the computational efficiency of WeiDetect, we compared the server-side execution time required for aggregation and malicious model detection across various defense mechanisms discussed in this paper. Table \ref{tab:computation_overhead} provides a detailed comparison of the computational overhead for each approach. To ensure a fair comparison, the computational cost for each approach was assessed under similar attack conditions.
\begin{table}[h!]
\centering

\caption{Comparison of WeiDetect's server-side FL aggregation computational time (in seconds) with state-of-the-art strategies.}
\label{tab:computation_overhead}
\begin{tabular}{lcc}
\hline
\multirow{2}{*}{Method$\downarrow$}&\multicolumn{2}{c}{Dataset}\\
\cmidrule(lr){2-3}
 &  CIC-Darknet2020 & CSE-CIC-IDS2018 \\ \hline
 Krum &0.4119&0.2979\\
 MKrum &1.0575  &0.9560\\
 TMean &0.0030&0.0020\\
Median &0.0019&0.0040\\
FLAME&0.0599&0.0239\\
WeiDetect&0.0825&0.2252\\

\hline
\end{tabular}

\end{table}
From the results, it is evident that MKrum exhibits the highest execution time for detecting malicious clients. In contrast, the proposed WeiDetect approach demonstrates significantly lower computational overhead for malicious client identification and aggregation; WeiDetect took approximately 12.82 times faster than MKrum. Methods like Mean and Median demonstrate lower execution times, ranging between\textbf{ 1 and 4ms}. This is because the distances between individual local updates become increasingly computationally intensive as the number of clients grows.

\subsection{Analysis of WeiDetect with different volumes of auxiliary dataset}
In this section, we analyze the performance of WeiDetect with varying volumes of auxiliary datasets. The auxiliary dataset comprises instances with a high probability of being correctly classified into their actual class and exhibits high prediction confidence from the model. For the experiments discussed in section \ref{sec:experimentalsection}, WeiDetect utilized the entire set of samples in the auxiliary dataset for its performance evaluation.

To investigate the performance of WeiDetect further, we repeat the experiments by simulating auxiliary datasets with varying sample volumes per class, assessing its ability to handle scenarios where the auxiliary data is not always optimal or imbalanced in real-world cases. First, we evaluate the performance of WeiDetect in an adversarial scenario where compromised participants poison 10\% of the samples within the targeted class. Different experiments were conducted to evaluate the effectiveness of WeiDetect against various adversarial approaches, including FGSM, PGD, and Gaussian Noise (GN). In these experimental scenarios, malicious participants focus on executing adversarial attacks on triple-targeted classes. Furthermore, to examine the effect of auxiliary dataset size on performance, we systematically varied the auxiliary data volume by considering different proportions, denoted as $AuxiliaryVol$. Specifically, we tested values with ${10, 20, 30, 40, 50, 60, 70, 80, 90, 100}$ values for $AuxiliaryVol$.
\begin{table*}[h]
\centering
\caption{Performance of WeiDetect across varying auxiliary dataset volumes where compromised participants poison 10\% and 30\% of the samples in the targeted class.}
\label{tab:analysisdiffvolume}
\footnotesize
\resizebox{\textwidth}{!}{
\renewcommand{\arraystretch}{1.1}
\begin{tabular}{ccccccccccccc}
\toprule
\multicolumn{13}{c}{CIC-Darknet2020} \\ 

\hline
Poison ratio $\rightarrow$ &\multicolumn{6}{c}{10\%} &\multicolumn{6}{c}{30\%} \\ 
\cmidrule(lr){1-7}
\cmidrule(lr){8-13}
\cmidrule(lr){2-7}
\cmidrule(lr){8-13}
& \multicolumn{2}{c}{FGSM} & \multicolumn{2}{c}{PGD} & \multicolumn{2}{c}{GN}& \multicolumn{2}{c}{FGSM} & \multicolumn{2}{c}{PGD} & \multicolumn{2}{c}{GN}\\
\cmidrule(lr){2-3}
\cmidrule(lr){4-5}
\cmidrule(lr){6-7}
\cmidrule(lr){8-9}
\cmidrule(lr){10-11}
\cmidrule(lr){12-13}
$AuxillaryVol\downarrow$&  F1-score& CTime &F1-score& CTime& F1-score& CTime &  F1-score& CTime &F1-score& CTime& F1-score& CTime\\
\hline

10\% &\cellcolor{gray!60}87.915&106.123 &87.565&82.509& 87.484&86.859&87.836 &82.135 &\cellcolor{gray!60}87.497&82.201 &\cellcolor{gray!60}87.070& 82.298\\
20\% &87.941&142.853 &\cellcolor{gray!60}87.781&123.910&87.698&133.303&87.885 &130.554 & 87.932 &125.376&86.246 &127.158\\
30\% &88.014&187.257&87.985 &173.206&\cellcolor{gray!60}87.774 &181.046&\cellcolor{gray!60}88.016 & 175.782 & 87.352 &154.869&86.918 &173.242\\
40\%&88.153&232.995&88.058 &215.740&87.834&233.012 & 88.100  &214.766& 88.017 &213.248&86.886 &216.476 \\
50\%  &88.203&341.816&87.595&257.517&87.732& 279.676 & 87.970  &256.810& 87.759 &258.369&86.864 &255.979\\
60\%  & 88.244&330.605&88.149&298.786&88.021&307.381&88.072  &298.752 & 87.985 &299.535&87.061 &296.419\\
70\%&88.093&376.818&88.181&347.518 &88.152& 343.084&88.134  &344.424 &88.077 &339.909&87.113& 337.645\\
80\% &88.034&421.982&88.261&388.507 &88.195& 393.351 & 88.016 &388.339 &87.899&389.797 &86.782&387.797 \\
90\% &88.426&466.273& 88.404&432.231 &87.997& 433.486& 87.990 &431.286 & 87.815 &434.055&86.874 &428.285\\
100\%&89.687&551.510& 88.472&501.487&89.373&489.758& 89.717 & 470.635&88.319 &490.986&89.395 &470.786\\

\bottomrule
\end{tabular}}
\footnotesize CTime: Computation time in seconds
\end{table*}

Table \ref{tab:analysisdiffvolume} presents the performance of WeiDetect with varying volumes of the auxiliary dataset, where $10\%$ of the samples are poisoned. We analyze the F1-score of the global model to determine the optimal value for this parameter and assess how WeiDetect performs under different volumes of $AuxiliaryVol$. The highlighted cell indicates the F1-score of WeiDetect, which is closest to the highest F1-score achieved by the state-of-the-art approaches (excluding WeiDetect), highlighting the minimal volume of the auxiliary dataset required for WeiDetect to outperform other methods. 
From the table, it is evident that for the CIC-Darknet2020 dataset, even when the $AuxiliaryVol$ is 10\%, the proposed approach outperforms the other methods in the case of the FGSM attack. However, for the PGD attack, 20\% of the auxiliary dataset is required to achieve the best performance, while for GN, 30\% is optimal. 
Additionally, we analyze the computational time required for the server to identify and remove malicious models and the aggregation process. 
As the number of samples in the auxiliary dataset increases, the execution time of WeiDetect also increases.

To identify key trends, we analyze scenarios where 30\% and 50\% of the samples are poisoned. In both cases, WeiDetect consistently outperforms all other state-of-the-art methods. With the auxiliary dataset, using at least 10\% effectively defends against PGD and GN attacks for 30\% and 50\% of the poisoned samples. However, for the FGSM attack, defending against 30\% poisoned samples requires 30\% of the auxiliary dataset, while 60\% is necessary for 50\% poisoned samples.
Appendix Table \ref{tab:darknetanalysisdiffvolume_50} presents additional results, reinforcing WeiDetect’s robustness across different poisoning scenarios. We also conduct this analysis on the CSE-CIC-IDS2018 dataset, as shown in Appendix Table \ref{tab:ids_analysisdiffvolume}. The results confirm that the proposed approach surpasses other methods, requiring at least 90\% auxiliary dataset samples to identify malicious models in this dataset effectively.



\section{Conclusion }
\label{sec:conclusion}
This paper introduces an FL-based NIDS leveraging the Weibull distribution with a validation score mechanism, WeiDetect. This approach addresses key challenges in FL-based NIDS, including non-IID data distributions and the risk of poisoning attacks.
First, we analyze the performance of an FL-based NIDS under poisoning attacks, where adversarial samples are generated using an adversarial sample generation approach. We then compare the results of our proposed defense mechanism against these attacks with state-of-the-art approaches. Our findings demonstrate that WeiDetect outperforms existing defense methods by achieving higher recall for the target class while maintaining the global model's F1 score. Secondly, we also evaluate the performance of WeiDetect under the label-flipping attack case.

In the future of this work, we aim to extend the research to other cyber threats, such as model inversion and inference attacks, and investigate the impact of federated unlearning in NIDS. Additionally, we plan to evaluate WeiDetect’s adaptability to other resource-constrained domains.

\section*{Acknowledgments}
\begin{figure}[h]
\centering
\includegraphics[width=0.5\linewidth]{ 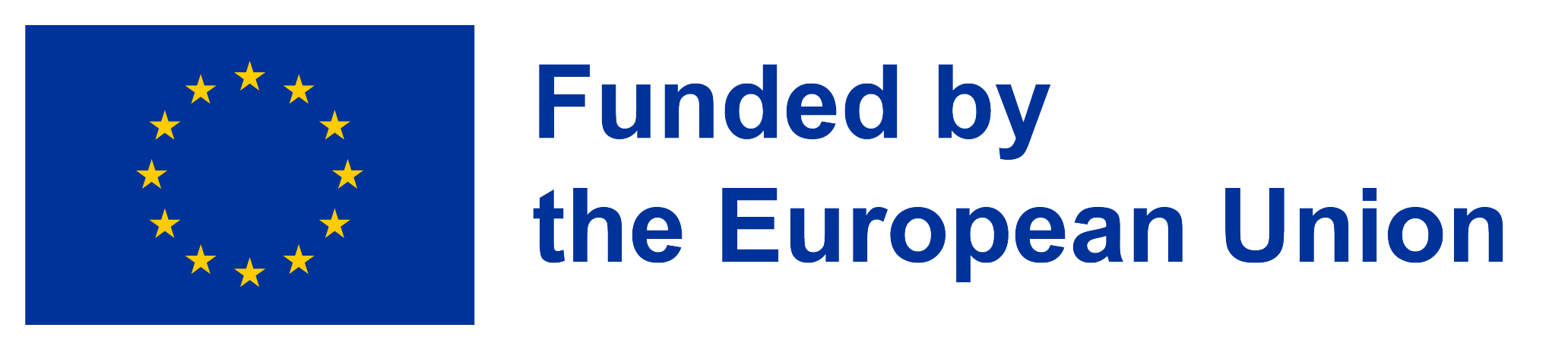} 
\end{figure}
This work was partly supported by the HORIZON Europe Framework Programme through the project ``OPTIMA- 
 Organization sPecific Threat Intelligence Mining and sharing" ( 101063107), funded by the European Union. Views and opinions expressed are however those of the author(s) only and do not necessarily reflect those of the European Union. Neither the European Union nor the granting authority can be held responsible for them. Moreover, the authors also thank the support of São Paulo Research Foundation (Fapesp) under the grant Horus \#2023/12865-8.  

\bibliographystyle{elsarticle-num} 
\bibliography{reference}

\section*{Appendix}
\appendix

\begin{table}[ht]
\centering
\caption{Performance of WeiDetect across varying auxiliary dataset volumes where compromised participants poison 50\% of the samples in the targeted class.}
\label{tab:darknetanalysisdiffvolume_50}
\footnotesize
\resizebox{0.45\textwidth}{!}{
\begin{tabular}{ccccccc}
\toprule
\multicolumn{7}{c}{CIC-Darknet2020} \\ 

\hline
Attack Strategy $\rightarrow$& \multicolumn{2}{c}{FGSM} & \multicolumn{2}{c}{PGD} & \multicolumn{2}{c}{GN}\\
\cmidrule(lr){1-1}
\cmidrule(lr){2-3}
\cmidrule(lr){4-5}
\cmidrule(lr){6-7}
$AuxillaryVol\downarrow$&  F1-score& CTime &F1-score& CTime& F1-score& CTime \\
\hline

10\% &87.480 &90.4247& \cellcolor{gray!60}86.843&92.074 &  \cellcolor{gray!60}86.865&92.074\\
20\% &87.409 &133.729&87.525 &150.933 &  87.056  &150.933 \\
30\% & 87.084 &180.834&87.615  &200.873&  87.337 & 200.873\\
40\% &87.072 &200.873& 87.789 &248.086& 87.397 &248.086 \\
50\% &88.228 &278.643&87.585 & 268.756&  87.416 & 268.756\\
60\% & \cellcolor{gray!60}87.711  &327.426&87.753  &307.523&  87.729 & 307.523\\
70\%&87.628 &374.753&88.105 & 341.419&  87.803  &341.419 \\
80\% &87.688 & 431.702&88.046 &391.741 & 87.866 &391.741\\
90\% &86.846 &477.629& 87.721 &440.772&  87.750 & 440.772\\
100\%&89.674 &525.241& 88.195 &475.542& 89.336  &475.542 \\
\bottomrule
\end{tabular}}
\\
\footnotesize CTime: Computation time in seconds
\end{table}

\begin{table*}[ht]
\centering
\caption{Performance of WeiDetect across varying auxiliary dataset volumes where compromised participants poison 10\% and 30\% of the samples in the targeted class in the CSE-CIC-IDS2018.}
\label{tab:ids_analysisdiffvolume}
\footnotesize
\resizebox{\textwidth}{!}{
\renewcommand{\arraystretch}{1.1}
\begin{tabular}{ccccccccccccc}
\toprule
\multicolumn{13}{c}{CSE-CIC-IDS2018} \\ 

\hline
Poison ratio $\rightarrow$ &\multicolumn{6}{c}{10\%} &\multicolumn{6}{c}{30\%} \\ 
\cmidrule(lr){1-1}
\cmidrule(lr){2-7}
\cmidrule(lr){8-13}
\cmidrule(lr){2-7}
\cmidrule(lr){8-13}
Approach $\rightarrow$ & \multicolumn{2}{c}{FGSM} & \multicolumn{2}{c}{PGD} & \multicolumn{2}{c}{GN}& \multicolumn{2}{c}{FGSM} & \multicolumn{2}{c}{PGD} & \multicolumn{2}{c}{GN}\\
\cmidrule(lr){2-3}
\cmidrule(lr){4-5}
\cmidrule(lr){6-7}
\cmidrule(lr){8-9}
\cmidrule(lr){10-11}
\cmidrule(lr){12-13}
$AuxillaryVol\downarrow$&  F1-score& CTime &F1-score& CTime& F1-score& CTime &  F1-score& CTime &F1-score& CTime& F1-score& CTime\\
\hline

10\% &92.816& 447.95&92.820 & 434.75 & 92.787&462.9&92.743& 430.15&92.809 &435.54&92.485 &428.65\\
20\% & 92.964&641.7&92.980 &630.45  & 92.895
&671.65&92.867&638.25 &92.928 &632.7& 92.631&625.75\\
30\% & 92.962&858.4&92.965 & 824.85 & 92.891&888.55&92.901& 829.5& 92.929&828.18&92.649 &821.59\\
40\% &92.963 &1060.05&92.954 &1027.5  & 92.778
&1084.95&92.912&1063.25 &92.932 &1174.12& 92.632&1038.05\\
50\% &92.805 &1239.85&92.793 &1202.28  & 92.893&1331.85&92.772& 1316.85& 92.823&1395.8& 92.889&1229.7\\
60\% & 92.964&1603.9 &92.965 & 1386.95 & 93.064&1527.45&92.959& 1517.27& 92.932&1413.9& 92.639&1425.4\\
70\%& 92.807&1748.28&92.797 &1602.45  & 92.893&1739.4&92.775&1747.1 & 92.823&1577.95&92.886 &1588.25\\
80\% &92.964 &1803.55& 92.953 & 1783.15 & 92.894
&1932.9&93.117&1954.15 & 93.264&1778.4& 93.025&1760.8\\
90\% &93.121 &1998.75&92.980 &1956.75  &93.208 &2139.7&93.00&2165.9 & 93.264&1941.95&93.026 &1964.75\\
100\% &93.285 &2126.25&93.295 &2333.80 &93.283 &2324.25&93.225&2346.75 &93.402 &2103&93.265 &2120.8\\
\bottomrule
\end{tabular}}
\footnotesize CTime: Computation time in seconds
\end{table*}

\end{document}